\documentclass[fleqn,usenatbib]{mnras}

\usepackage[english]{babel}
\usepackage{amsmath,amssymb}
\usepackage{graphicx, subfig}
\usepackage[normalem]{ulem}

\usepackage{verbatim}

\usepackage{color}
\usepackage{multirow}
\usepackage{mathtools}
\usepackage{epstopdf}

\usepackage{url}
\usepackage{xcolor}
\usepackage{orcidlink}

\AtBeginShipout{%
  \ifnum\value{page}>1 %
    \typeout{* Additional boxing of page `\thepage'}%
    \setbox\AtBeginShipoutBox=\hbox{\copy\AtBeginShipoutBox}%
  \fi
}

\newcommand\code[1]{\textsc{\MakeLowercase{#1}}}
\newcommand{\quotes}[1]{``#1''}

\newcommand{\footnoteref}[1]{\textsuperscript{\ref{#1}}}

\def\msun{{\rm M}_{\odot}}
\def\zsun{{\rm Z}_{\odot}}
\def\lsun{{\rm L}_{\odot}}
\def\dsun{\mathcal{D}_{\odot}}
\def\gnot{{\rm G}_{0}}

\def\pc{{\rm {pc}}}
\def\kpc{{\rm {kpc}}} 
\def\mpc{{\rm {Mpc}}} 
 
\def\angstrom{\textrm{A\kern -1.3ex\raisebox{0.6ex}{$^\circ$}}}

\def\cc{{\rm cm}^{-3}}
\def\myr{{\rm Myr}}
\def\gyr{{\rm Gyr}} 
\def\invgyr{{\rm Gyr}^{-1}} 
\def\kms{\,{\rm {km\, s^{-1}}}} 

\def\msunyr{\msun\,{\rm yr}^{-1}}
\def\surfd{\msun\,{\rm kpc}^{-2}}
\def\surfsfr{\msun\,{\rm yr}^{-1}\,{\rm kpc}^{-2}}

\def\surfl{\lsun\,{\rm kpc}^{-2}}

\def\SFR{{\rm SFR}}
\def\sSFR{{\rm sSFR}}
\def\sigmasfr{\Sigma_{\rm SFR}}

\def\HH{${\rm {H_2}}$}

\def\OIIIion{${\rm O}^{++}$}

\def\nHH{n_{\rm H2}}

\def\dust{\mathcal{D}}

\def\CII{\hbox{[C~$\scriptstyle\rm II $]}}
\def\OI{\hbox{[O~$\scriptstyle\rm I $]}} 
\def\OIII{\hbox{[O~$\scriptstyle\rm III $]}} 

\def\CIII{\hbox{C~$\scriptstyle\rm III $]}}

\def\mhalo{M_{\rm h}}
\def\rvir{r_{\rm vir}}

\def\be{\begin{equation}} 
\def\ee{\end{equation}} 
\def\ba{\begin{eqnarray}} 
\def\ea{\end{eqnarray}} 
\def\gtsima{$\; \buildrel > \over \sim \;$}
\def\ltsima{$\; \buildrel < \over \sim \;$}
\def\gsim{\lower.5ex\hbox{\gtsima}} 
\def\lsim{\lower.5ex\hbox{\ltsima}}
\def\prosima{$\; \buildrel \propto \over \sim \;$} 
\def\simgt{\lower.5ex\hbox{\gtsima}} 
\def\simlt{\lower.5ex\hbox{\ltsima}} 
\def\simpr{\lower.5ex\hbox{\prosima}}

\definecolor{mkcolor}{HTML}{01abdf}
\definecolor{apcolor}{HTML}{b3003b}
\definecolor{afcolor}{HTML}{01bdff}
\definecolor{lvcolor}{HTML}{ff9933}

\def\nat{Nature}
\def\na{New Astronomy}

\def\mnras{MNRAS}
\def\apj{ApJ}
\def\apjl{ApJL}
\def\apjs{ApJS}
\def\aap{A\&A}
\def\aaps{A\&A Supp.}
\def\aapr{A\&A Rev.}
\def\araa{ARA\&A}
\def\aj{AJ}

\def\physrep{Phys. Rep.}
\def\pasa{Publ. Astr. Soc. Australia}
\def\pasp{Publ. Astr. Soc. Pac.}

\def\pasj{Pub. Astron. Soc. Japan}

\graphicspath{{plots_png/},{plots/}}

\title[A SERRA of galaxies]{A survey of high-$z$ galaxies: SERRA simulations}
\author[A. Pallottini et al.]{A. Pallottini \orcidlink{0000-0002-7129-5761}$^{1}$\thanks{\href{mailto:andrea.pallottini@sns.it}{andrea.pallottini@sns.it}},
 A. Ferrara\orcidlink{0000-0002-9400-7312}$^{1}$,
 S. Gallerani$^{1}$,
 C. Behrens$^{2}$,
 M. Kohandel\orcidlink{0000-0003-1041-7865}$^{1}$,
 S. Carniani\orcidlink{0000-0002-6719-380X}$^{1}$, 
 L. Vallini\orcidlink{0000-0002-3258-3672}$^{1}$, \newauthor
 S. Salvadori\orcidlink{0000-0001-7298-2478}$^{3,4}$,
 V. Gelli\orcidlink{0000-0001-5487-0392}$^{3,4}$,
 L. Sommovigo$^{1}$,
 V. D'Odorico\orcidlink{0000-0003-3693-3091}$^{1,5,6}$,
 F. Di Mascia$^{1}$,
 E. Pizzati\orcidlink{0000-0002-9712-0038}$^{1,7}$.
\\
$^{1}$ Scuola Normale Superiore, Piazza dei Cavalieri 7, I-56126 Pisa, Italy\\
$^{2}$ Institut f\"{u}r Astrophysik, Georg-August Universit\"{a}t G\"{o}ttingen, Friedrich-Hund-Platz 1, 37077, G\"{o}ttingen, Germany\\
$^{3}$ Dipartimento di Fisica e Astronomia, Universitá degli Studi di Firenze, via G. Sansone 1, 50019, Sesto Fiorentino, Italy\\
$^{4}$ INAF - Osservatorio Astrofisico di Arcetri, Largo E. Fermi 5, I-50125, Firenze, Italy\\
$^{5}$ INAF - Osservatorio Astronomico di Trieste, via G.B. Tiepolo, 11 I-34143 Trieste\\
$^{6}$ Institute for Fundamental Physics of the Universe, Via Beirut 2, I-34151 Miramare, Trieste, Italy\\
$^{7}$ Leiden Observatory, Leiden University, Niels Bohrweg 2, NL-2333 CA Leiden, the Netherlands\\
}

\date{Accepted XXX. Received XXX; in original form XXX}

\pubyear{2022}

\begin{document}
\label{firstpage}
\pagerange{\pageref{firstpage}--\pageref{lastpage}}
\maketitle
%
\begin{abstract}
We introduce \code{SERRA}, a suite of zoom-in high-resolution ($1.2\times 10^4 \msun$, $\simeq 25\,\pc$ at $z=7.7$) cosmological simulations including non-equilibrium chemistry and on-the-fly radiative transfer. The outputs are post-processed to derive galaxy UV+FIR continuum and emission line properties. Results are compared with available multi-wavelength data to constrain the physical properties (e.g., star formation rates, stellar/gas/dust mass, metallicity) of high-redshift $6 \lsim z \lsim 15$ galaxies. This flagship paper focuses on the $z=7.7$ sub-sample, including 202 galaxies with stellar mass $10^7 \msun \lsim M_\star \lsim 5\times 10^{10}\msun$, and specific star formation ranging from ${\rm sSFR} \sim 100\,\gyr^{-1}$ in young, low-mass galaxies to $\sim 10\,\gyr^{-1}$ for older, massive ones. At this redshift, \code{serra} galaxies are typically {bursty}, i.e. they are located above the Schmidt-Kennicutt relation by a factor $\kappa_s = 3.03^{+4.9}_{-1.8}$, consistent with recent findings for \OIII~and \CII~emitters at high-$z$. They also show relatively large IRX $=L_{\rm FIR}/L_{\rm UV}$ values as a result of their compact/clumpy morphology effectively blocking the stellar UV luminosity. Note that this conclusion might be affected by insufficient spatial resolution at the molecular cloud level. We confirm that early galaxies lie on the standard \CII$-\rm SFR$ relation; their observed $L_{\rm [OIII]}/L_{\rm [CII]} \simeq 1-10$ ratios can be reproduced by a part of the \code{SERRA} galaxies without the need of a top-heavy IMF and/or anomalous C/O abundances. \OI~line intensities are similar to local ones, making ALMA high-$z$ detections challenging but feasible ($\sim 6\,\rm hr$ for an SFR of $50\,\msunyr$).
\end{abstract}


\section{Introduction}

Understanding the properties of galaxies in the Epoch of the Reionization (EoR, redshift $z \gsim 5.5$) is an urgent quest in modern physical cosmology. In the last decades, optical/near infrared (IR) surveys have been used to identify galaxies, measure their star formation rates, and follow their stellar mass build-up up to $z \sim 10$ \citep{dunlop:2013,bouwens:2015,oesch:2018}. These data, in turn, inform cosmic reionization and metal enrichment models \citep[see][for recent reviews]{dayal:2018,maiolino:2018}.

Despite such magnificent progress, we know very little about a key component of the galactic ecosystem, i.e. the interstellar medium (ISM) of the early, \quotes{normal}, star-forming (${\rm SFR}\lsim 100\,\msunyr$) galaxy population. 
While James Webb Space Telescope (JWST) is about to give us access to the strong emission lines typically used to characterise the ISM properties at low-$z$ \citep[e.g.][]{kewley:2002,curti:2017}, possibly augmented by machine learning analysis techniques \citep{ucci:2019}, a complementary approach consists of using Far Infrared (FIR) lines.

The Atacama Large Millimeter/submillimeter Array (ALMA) has enabled such exploration, in particular by exploiting the prominent \citep{stacey:1991} \CII~$158\mu$m line. The earliest investigations of $z>5$ targets reported a mix of upper limits \citep{ouchi:2013,kanekar:2013,ota:2014,schaerer:2015} and detections \citep{capak:2015,maiolino:2015,watson:2015,willott:2015}.
After such pioneering stage, \CII~is now routinely observed both in Lyman Break Galaxy \citep[LBG,][]{carniani:2017,carniani:2018,smit:2018,hashimoto:2019,bakx:2020}, Lyman Alpha Emitters \citep[LAE,][]{matthee:2017,carniani:2018_b,harikane:2018}, and in gravitationally lensed, low mass objects \citep{pentericci:2016,knudsen:2016,bradav_c:2017}.
Today, FIR surveys containing hundreds of galaxies are being produced and analysed, both at just after \citep[$4\lsim z \lsim 6$][\code{ALPINE}]{le_fevre:2019} and well within \citep[$6\lsim z \lsim 8$][\code{REBELS}]{bouwens:2021} the EoR.

Locally, it has been shown \citep[e.g.][]{cormier:2015} that \CII~traces essentially all ISM phases (cold/warm neutral, ionized) thus carrying very valuable and diverse information \citep[see also][]{olsen:2018}. Initially, the comparison between EoR \citep{carniani:2018,schaerer:2020} and local \citep{de_looze:2014,herrera-camus:2015,herrera-camus:2018} data has mostly concentrated on the \CII-SFR relation. Theoretical works followed the same path, by focusing on the interpretation of such relation and its connection with the burstiness of the star formation process \citep{ferrara:2019,pallottini:2019,vallini:2020}, parameterized via a deviation from the locally observed Schmidt-Kennicutt \citep[][]{schmidt:1959,kennicutt:1998} relation, and/or other parameters such as the ISM metallicity, density and interstellar radiation field \citep{vallini:2013,vallini:2015,katz:2017,olsen:2017,pallottini:2017,pallottini:2017_b,popping:2019}.

The investigation of early galaxies via \CII~line has nevertheless brought some genuine surprises, the most conspicuous of which is the detection of extended (up to 10 kpc) \CII~halos around these sources. Such halos have been first discovered in stacking experiments \citep{fujimoto:2019,ginolfi:2020}, and later confirmed individually in the most massive galaxies of the \code{ALPINE} sample \citep{fujimoto:2020}. A tantalising explanation of this feature is that \CII~halos are created by supernovae- and/or AGN-driven outflows \citep{pizzati:2020} transporting carbon and other heavy elements in the circumgalactic medium. Notably, the \CII~line in these sources often shows extended wings produced by fast outflowing gas \citep{gallerani:2018, ginolfi:2020}.

We finally mention that \CII~observations are becoming deep enough to perform kinematic analysis \citep{jones:2017,smit:2018,schaerer:2020}, and interpreting via sophisticated models \citep{kohandel:2019,kohandel:2020}: accurate rotation curves might allow us to set constraints on the role and amount of dark matter in these objects.

Given the multi-component origin of the \CII~emission, complementary probes arising from a specific ISM phase are very valuable. Fortunately, the \OIII~$88\mu$m, uniquely tracing dense ionised regions of the ISM, is observable and comes into rescue. Searches for such line \citep{inoue:2016_b,carniani:2017,laporte:2017,tamura:2018,hashimoto:2019,carniani:2020} have motivated interpretations of the observed morphological differences with respect to \CII~\citep{pallottini:2019,carniani:2020,harikane:2020,arata:2020}, and its use as a proxy for metallicity \citep{olsen:2015,inoue:2016_b,vallini:2017}.

The molecular gas content and InterStellar Radiation Field (ISRF) can instead be probed via CO~roto-vibrational lines \citep{vallini:2018,vallini:2019}. High-$J$ CO lines can also be used to explore the possible presence of X-ray sources and shocks in these galaxies \citep{Gallerani14,vallini:2019}.
However, CO emission is faint and it has only been detected in few EoR objects \citep{pavesi:2019,dodorico:2018}, sometimes in a serendipitous manner
\citep{feruglio:2018}.

As for the dust FIR continuum emission, it entails considerable uncertainties as the dust mass and temperature are degenerate and often hard to determine in high-$z$ sources \citep[e.g.][]{fudamoto:2020,faisst:2020}. Available ALMA observations sample the FIR regime with very few data points (sometimes even a single one) making it hard to break the degeneracy. Some theoretical works predict that dust should be warmer at high-$z$ \citep{behrens:2018,liang:2019,sommovigo:2020,sommovigo:2021,dimascia:2021}; others claim that dust production is more efficient early on \citep{mancini:2015,popping:2017}.
To make progress, it would be necessary to probe wavelengths blue-ward of the black-body peak by means of mid-infrared observations. Given the cancellation of the SPICA \citep[Space Infrared Telescope for Cosmology and Astrophysics][]{spinoglio:2017,egami:2018} and OST \citep[Origins Space Telescope][]{originswhitepaper2020} missions, novel strategies must be urgently devised.

To conclude, the study of the ISM of EoR galaxies is in its infancy. The investigation of the ISM is crucial to answering key open physical questions concerning these sources: what is their star formation efficiency? how various (mechanical, radiative, chemical) feedback processes affect the thermal, dynamical, and turbulent state of the ISM? what is the amount of dust, metals, and molecular gas in these systems? These questions could be finally attacked by combining ALMA and JWST data.

From a theoretical point of view, progressively more refined and physics-rich numerical simulations represent an indispensable guideline and tool. 
Since we would like to resolve the ISM down to the Molecular Clouds (MC) level, suitable simulations should reach a (baryonic) mass resolution $\simeq 10^4 \msun$, and be able to probe $\simeq 10\, \rm pc$ scales. While not as critical as for simulations that aim at studying the reionization process (e.g. \code{CROC} \citealt{gnedin:2014}, \code{CODA II} \citealt{ovrick:2018}, \code{SPHINX} \citealt{rosdahl:2018}), the ISRF is an important physical ingredient for early galaxy evolution \citep{katz:2017,trebitsch:2017,pallottini:2019}. Thus, radiative transfer should be consistently implemented.

As smaller scales are resolved, a better prescription of unresolved (sub-grid) processes must be included with respect to coarser large-scale cosmological simulations (mass resolution $\simeq 10^6 \msun$ and $\simeq 0.2-1\, \rm kpc$ scales, e.g. \code{EAGLE} \citealt{schaller:2015}, \code{BlueTides} \citep{feng:2016}, \code{Illustris-TNG} \citealt{pillepich:2018}, and \code{THESAN} \citealt{kannan:2021}), in order to properly account for the turbulent structure of the ISM, its chemical state, and stellar feedback \citep[][]{agertz:2013,agertz:2015,hopkins:2014,pallottini:2017,pallottini:2017_b,lupi:2019}.
This makes it demanding to simulate large galaxy sample in an unbiased cosmological volume with high resolution and a variety of sub-grid models\footnote{See, e.g. the \quotes{normal} volume of the \code{Renaissance} from \citet[][]{o_shea:2015}, where the simulation focuses on a $\simeq (6 \,\mpc^3)$ region and it is stopped around $z \simeq 12$.}.

A more efficient option is to perform multiple zoom-in simulations, as done e.g. in \citet[][see \code{FIRE-2} \citealt{hopkins:2018} for the models]{ma:2018}, \citet[][\code{FirstLight}]{ceverino:2017}, and \citet[][\code{FLARES}]{lovell:2021}.
With some efforts \citep[see in particular][]{lovell:2021}, multiple zoom-in simulations can be used to construct a composite sample that is representative of the parent cosmological volume used for the zoom-in.

In this paper we present \code{Serra}\footnote{\code{Serra} means greenhouse in Italian. You might notice that galaxies are named after flowers (and some also look alike).}, a suite of multiple zoom-in cosmological simulations targeting $M_\star \sim 10^8 - 10^{10} \msun$ galaxies at $z=6$. Each simulation follows the evolution of $10-30$ galaxies in a cosmological environment by employing a radiation-hydrodynamic code that is coupled to a non-equilibrium chemical network. In the present work, we analyse the results from 10 simulations focusing at $z=7.7$, for a total of 202 galaxies.
Here, we do not aim at building a fully unbiased cosmological galaxy sample. Rather, we intend to mimic a survey targeting galaxies that are luminous enough to be detected and characterised with current (Hubble Space Telescope and ALMA) and upcoming instruments (JWST). Paralleling current \CII~observational campaigns with ALMA \citep[][]{le_fevre:2019,bouwens:2021}, we can then reconstruct at a later stage the statistical properties \citep{bethermin:2020} of our sample, at the same time spotting interesting or peculiar objects \citep{loiacono:2020,fudamoto:2021}. 

Most of the assumptions and techniques adopted in \code{SERRA} have already been developed in previous works. In \citet[][]{pallottini:2017} we focused on the modelling of stellar feedback; \citet[][]{pallottini:2017_b} showed the impact of non-equilibrium chemistry; in \citep[][]{pallottini:2019} we analysed the impact of radiation-hydrodynamical effects. Advanced line and continuum emission models, a foundational feature of \code{serra}, are based on \citet{vallini:2015,vallini:2019} and \citet{behrens:2018}, respectively. The simulation results are post-processed to produce synthetic data cubes \citep{kohandel:2020}, which can be directly and fairly compared with observations \citep[][]{behrens:2018,zanella:2021,gelli:2021}.

The paper plan is as follows. In Sec. \ref{sec_models}, We summarise the main features and assumptions of the \code{Serra} simulations.
In Sec. \ref{sec_results}, we present our results. We analyse the stellar mass build up and time evolution of the simulated galaxies (Sec.s \ref{sec_analysis_evolution} and \ref{sec_main_sequence}).
Focusing on the $z=7.7$ sample, we study the halo-mass stellar mass relation (Sec.s \ref{sec_analysis_z_7_7} and \ref{Sec:Alyssum}), the Schmidt-Kennicutt relation (Sec. \ref{sec_sk_relation}), and the size-mass relation (Sec. \ref{sec_size_stellar_mass})
Then (Sec. \ref{sec_emission_7_7}), we explore the main FIR and UV properties, treating both continuum (Sec.s \ref{sec_uv_production} and \ref{Sec:continuum}) and line (Sec. \ref{sec_line_emission}) emission.
Sec. \ref{sec_conclusions} provides a summary.

\section{Numerical simulations}\label{sec_models}

In \code{SERRA} the dynamical evolution of dark matter (DM), gas, and stars is simulated with a customised version of the Adaptive Mesh Refinement (AMR) code \code{ramses}\footnote{\url{https://bitbucket.org/rteyssie/ramses}\label{foot_ramses}} \citep{teyssier:2002}. \code{ramses} follows the gas evolution with a second-order Godunov scheme; DM and stars are treated with a multigrid particle-mesh solver \citep[][]{guillet:2011}.
Radiative transfer is solved on-the-fly using \code{ramses-rt} \citep{rosdahl:2013}, which implements photon advection using a momentum-based framework with M1 closure for the Eddington tensor \citep{aubert:2008}.
While photon transport is handled by \code{ramses-rt}, the system of equations regulating gas-photon interactions (including the gas thermal evolution) is generated via \code{krome}\footnote{\url{https://bitbucket.org/tgrassi/krome}\label{foot_krome}} \citep{grassi:2014} adopting a non-equilibrium chemical network \citep{bovino:2016,pallottini:2017_b}. The coupling between \code{krome} and \code{ramses-RT} is done by sub-cycling the absorption step in order to assure convergence \citep[see also][]{decataldo:2019,pallottini:2019}.

Critical to a fair comparison with observations is the production of line and continuum emission. This is done in post-processing by adopting multiple models.
Emission lines are calculated by interpolating grids of models obtained from the photo-ionization code \code{cloudy} c17\footnote{\url{https://www.nublado.org}} \citep{ferland:2017}, by extending \citet{vallini:2017} models as described in \citet{pallottini:2019}.
Attenuated UV and continuum FIR emission is computed via the Monte Carlo code \code{SKIRT} V8\footnote{\url{https://www.skirt.ugent.be}} \citep{baes:2015,camps:2015}, using the modelling from \citet{behrens:2018}.

In the following, we summarize the main features of the numerical set-up (Sec. \ref{sec_ic}), radiation hydrodynamics (Sec. \ref{sec_rad_hydro}), stellar modelling (Sec. \ref{sec_stars}), galaxy identification and emission properties (Sec. \ref{sec_finders_emission}).

\subsection{Set-up}\label{sec_ic}

We generate cosmological initial conditions (IC) assuming parameters compatible with \emph{Planck} results\footnote{$\Lambda$CDM model with vacuum, matter, and baryon densities in units of the critical density $\Omega_{\Lambda}= 0.692$, $\Omega_{m}= 0.308$, $\Omega_{b}= 0.0481$, Hubble constant $\rm H_0=67.8\, {\rm km}\,{\rm s}^{-1}\,{\rm Mpc}^{-1}$, spectral index $n=0.967$, and $\sigma_{8}=0.826$ \citep[][]{planck_collaboration:2014}.}. IC are computed with \code{music} \citep{hahn:2011} at $z=100$ for a cosmological volume of $(20\,\mpc/ h)^{3}$, which will contain $\simeq 7$ halos with masses $\mhalo \geq 10^{11} \msun$ at $z=6$ \citep[][adopting a \citealt{sheth:1999} halo mass function]{murray:2013}. The base grid has 8 levels, equivalent to a gas (DM) mass of $6\times 10^6\msun$ ($3.6\times 10^{7}\msun$). We run a total of 11 simulations and for each simulation we prepare distinct IC; we then evolve the volume down to $z=6$. We use \code{hop} \citep{eisenstein:1998} to extract halo catalogues and select a target among the most massive halos in the simulated cosmological volume. These have $\mhalo \simeq (1-5)\times 10^{11} \msun$ (corresponding to $\simeq 3.2-4.1$ sigma fluctuations), and thus they are resolved with about $(2 - 16)\times 10^3$ particles in the coarse runs.

Using \code{hop} we select the minimum ellipsoid enveloping about $6$ times the halo virial radius ($\rvir$), and trace it back to $z=100$. Halos with $\mhalo = (1-5)\times 10^{11} \msun$ have virial radii of $\simeq 20 - 35 \,\kpc$. A comoving {\it trace back} radius of about $1 -2\,\mpc/h$ is typically obtained. We select such Lagrangian volume at $z=100$ and using \code{music} we produce multimass IC, i.e. we progressively add concentric layers of increasing mass resolution around the target. We reach 11 levels in the zoom-in region, i.e. a gas (DM) mass of $1.2\times 10^4 \msun$ ($7.2\times 10^4 \msun$). Such Lagrangian volume is then evolved at $z=6$ and the main halo is resolved with about $(1-7)\times 10^6$ DM particles in these zoom-in runs.

In the zoom-in region we set 6 additional AMR levels (thus 17 in total), that are enabled via a Lagrangian-like criterion. This allows us to reach spatial scales of $\simeq 30\,{\rm pc}$ ($\simeq 25\,{\rm pc}$) at $z=6$ ($z=7.7$) in the densest regions, i.e. the most refined cells have mass and size typical of Galactic MCs \citep[$10-50\,{\rm pc}$, $10^{4-6}\msun$][]{federrath:2013}.

Adopting a trace back radius of $\simeq 6\rvir$ typically ensures negligible contamination for the central halo \citep[][]{onorbe:2014}. However, here we are also interested in other galaxies in the proximity of the target. For this reason we perform the target selection and trace back procedure two times \citep[cfr. with][]{fiacconi:2015}. This limits the contamination levels to $\simeq 0.1\%$ ($\simeq 2\%$) for the target (surrounding) halos in the zoom-in region; from the analysis we exclude galaxies with a contamination of $\geq 4.3\%$ within $2\,\rvir$.
With this threshold, the resulting sample has an average\footnote{When giving averages, we are quoting the median value of the distribution with errors based on the values of $25\%$ and $75\%$ of the dataset.} contamination level of $0.0^{+0.7}_{-0.0} \%$ ($1.1^{+0.6}_{-0.4} \%$) when considering $1\,\rvir$ ($2\,\rvir$).
Further, the apparently higher contamination of some of the galaxies in the simulations \citep[][]{onorbe:2014} is mostly due to the adopted definition, i.e. we are considering all DM particles within $2\,\rvir$ (of the host DM) from the centre of a galaxy, while $1\,\rvir$ is adopted in most works.

\subsection{Radiation hydrodynamics}\label{sec_rad_hydro}

\subsubsection{Gas and dust composition}

The chemical network in the simulations includes H, H$^{+}$, H$^{-}$, He, He$^{+}$, He$^{++}$, H$_{2}$, H$_{2}^{+}$ and electrons. Each of these species is tracked individually, and evolves according to the chemical processes described in Sec. \ref{sec_chem_processes}. Metallicity ($Z$) is tracked as the sum of heavy elements, and we assume solar abundance ratios of the different metal species \citep{asplund:2009}. Numerically, metallicity is treated as a passively advected scalar; and stars act as a source of metals and reprocessed material (Sec. \ref{sec_stars}). Dust evolution is not explicitly tracked during simulation. We make the assumption that the dust-to-gas mass ratio scales with metallicity, i.e. $\dust = \dsun (Z/\zsun)$, where $f_d=\dsun/\zsun = 0.3$ for the Milky Way \citep[MW, e.g.][]{hirashita:2002}. For chemical reactions involving dust and radiation absorption, we assume a MW-like grain size distribution \citep{weingartner:2001}.

Individual ICs for the various chemical species are computed accounting for the chemistry in a primordial Universe \citep{galli:1998}. We adopt an initial metallicity floor $Z_{\rm floor}=10^{-3}\zsun$ since at $z \gsim 40$ our resolution does not allow us to track the formation of first stars \citep[see][]{wise:2012_b,oshea:2015,smith:2018}. Such floor only marginally affects the gas cooling time (see also Sec. \ref{sec_chem_processes}), and it is compatible with the metallicity of the intergalactic medium (IGM) in cosmological simulations \citep[e.g.][]{pallottini:2014,maio:2015,jaacks:2018}. Such setup is fully described in \citet{pallottini:2017_b}.

\subsubsection{Photons}\label{sec_radiation}

In \code{ramses-rt}, photons are treated as a fluid sharing the same AMR structure of the gas. Photons are divided into energy bins, whose density and velocity fields are tracked separately.

As done in \citep{pallottini:2019}, in \code{serra}, we follow 5 photon bins in the $6.0-24.59\,\rm eV$ range\footnote{For He and He$^+$ we assume collisional ionisation only.}.
The first two energy bins cover the \citet[][]{habing:1968} band\footnote{In this paper, the Habing flux $G$ is indicated in unit of $G_{0}=1.6\times 10^{-3} {\rm erg}\,{\rm cm}^{-2}\,{\rm s}^{-1}$, the MW value.} $6.0-13.6\,\rm eV$, which regulates the temperature of the ISM and photo-dissociation regions (PDR); of the two, the higher energy bin is specific for the Lyman-Werner radiation ($11.2-13.6\,\rm eV$), which photo-dissociates \HH~via the two-step Solomon process \citep{stecher:1967}.
The last three energy bins cover the H-ionising photons up to the first ionisation level of He ($13.6-24.59\,\rm eV$). For H-ionising photons, the energy width is chosen such that the bins have the same number of photons when the spectral energy distribution (SED) is averaged over a selected fiducial stellar population -- i.e. age $10\,\myr$ and $Z_{\star} = \zsun$ assuming \citet{bruzual:2003} photoionization models -- since these stars dominate the spectrum of a galaxy with an exponentially rising SFR \citep[e.g.][]{pallottini:2017_b}.

Gas and dust represent sinks for the radiation (Sec. \ref{sec_chem_processes}) while stars act as sources (Sec. \ref{sec_stellar_feedback}).
We neglect the cosmic UV background (UVB) contribution, since the typical ISM densities are sufficiently large to ensure an efficient self-shielding \citep[][]{gnedin:2010}. In the EoR, the UVB produces a statistically negligible hydrogen ionisation for gas with density $n \gsim 10^{-2}\cc$, i.e. close to those found in the outskirts of early galaxies \citep{rahmati:2013,chardin:2018}.
We do not account for recombination radiation, i.e. we assume that recombination photons are absorbed \quotes{on the spot} \citep{rosdahl:2013}.

Both \code{ramses} and \code{ramses-rt} have explicit schemes, thus the time step of a fluid with velocity $v$ is limited by the Courant condition \citep{courant:1928}, i.e. $\Delta t \sim \Delta l/v$. As radiation and gas evolution are coupled, the maximum speed is selected, i.e. the speed of light. To limit the computational load, we consider the \textit{reduced speed of light} approximation to propagate wave-fronts \citep{gnedin:2001}.

The reduced speed of light approach might produce artefacts in regions where the dissociation/ionisation front propagation is faster than the reduced speed of light, $c_{red}$, used in the simulation.
For instance, in reionization studies $c_{red}\sim 10^{-1} c$ allows to resolve the cosmological I-front propagation, yielding little variations on global reionization histories (\citealt{gnedin:2016}, see also \citealt{deparis:2018}, where the effect is explored in the $10^{-2}\leq c_{red}/c\leq 1$ range). For PDR regions, the speed of the wave-front can be estimated as
\be \label{eq:speed_front}
v_{\rm PDR} = \frac{G}{n \langle{\rm h }\nu\rangle_{\rm FUV}} \sim 5\times 10^{-3} \frac{G}{10^3 G_0} \frac{5\times 10^2 \cc}{n} c\,,
\ee
where we have used the typical values expected for high-$z$ galaxies where the photon-to-baryon ratio is particularly high, e.g. as in the case of a $\sim 10^5 \msun$ newly born star cluster in a typical GMC.

In \citet{pallottini:2019}, we adopted a conservative value of $c_{red} = 10^{-2} c$; however, there we also noted that the PDR region size is set by the dust optical depth $\tau_d=1$. This yields
\be
l_{\rm PDR} = \frac{\mathcal{D}}{\dsun} \frac{10^{21}\rm cm^{-2}}{n} \sim 0.6 \frac{\mathcal{D}}{\dsun} \frac{5\times 10^2 \cc}{n} \rm pc\,,
\ee
implying that PDR are typically unresolved in our simulations. The PDR region is established in a time
\be\label{eq:pdr_time}
t_{\rm PDR} = \frac{l_{\rm PDR}}{v_{\rm PDR}} \simeq 390 \frac{\mathcal{D}}{\dsun} \frac{10^3 G_0}{G} \rm yr\,
\ee
then the propagation of the front slow down below $v_{\rm PDR}$. Depending on the dust amount and the ISRF value, reducing the speed of light to a value lower then $c_{red} = 5\times 10^{-3} c$ is expected to affect only the evolution in a fraction of $\rm Myr$, as we also show in \citet[][]{decataldo:2019}.
A different situation is when radiation breaks out of the birth clouds and impacts the rest of the galaxy; after $\simeq 20 \,\myr$ a $\sim 10^5 \msun$ cluster would produce in its surrounding a ISRF of $G\sim 10\,\rm G_0$; assuming the SN feedback has cleared out the gas, we expect gas densities of the order of $n\sim 10^{-2}\,\cc$, thus from eq. \ref{eq:speed_front}  it follows that $v_{\rm PDR} = c$, i.e. the reduced speed of light approximation breaks down \citep[see][for a similar issue]{gnedin:2016}; however, the PDR region is already established in $\lsim 1\,\myr$ (eq. \ref{eq:pdr_time}); thus, the situation is expected to be relatively rare in our simulations\footnote{A similar reasoning applies also to H ionizing photons, since they are scarcely emitted after the stars producing them turn into SNs.}.
Moreover, the breakdown of the approximation is not expected to affect satellites, as the dominant contribution to the ISRF is generated in situ, because of geometrical dilution \citep{gelli:2022}.

Further, in \citet{lupi:2020} we adopted the same initial conditions used in \citet{pallottini:2019}, but with a different numerical code, and a lower speed of light, i.e. $c_{red} = 10^{-3} c$. We noted no large differences in properties of the target galaxy in the two zoom-in (e.g. SFR, $M_\star$, average ISM density, ISRF value, etc). This is consistent with what found in \citet{hopkins:2020}, where variation of $10^{-3}\lsim c_{red}\lsim 10^{2}$ are shown to affect the ISM properties trough a slight variation of the radiation pressure efficiency.

For this reason, we adopt $c_{red} = 10^{-3} c$ in the present, which -- with respect to \citet{pallottini:2019} -- also allow us to zoom-in relatively larger Lagrangian volumes and higher sigma fluctuations by spending roughly the same computational time.

\subsubsection{Chemical processes}\label{sec_chem_processes}

Our \code{krome}-powered non-equilibrium network follows a total of 48 reactions\footnote{The reactions, their rates, and corresponding references are listed in App. B of \citet{bovino:2016}: we use reactions from 1 to 31 and 53, 54, from 58 to 61, and from P1 to P9; the rates are reported in Tab. B.1, Tab. B.2, and Tab. 2 of \citet{bovino:2016}, respectively.}, including photo-chemistry, dust processes, and cosmic-ray-induced reactions \citep[][see \citealt{bovino:2016} for the original implementation]{pallottini:2017_b}.

Molecular hydrogen is particularly important in \code{serra}, since it determines the star formation process (Sec. \ref{sec_formation}). Formation of H$_{2}$ is possible both in gas-phase and on dust grain surfaces. The formation rate of \HH~on dust grains is approximated following \citet{jura:1975}:
\be\label{eq_jura}
R_{\rm H2-dust} = 3\times 10^{-17}n\,n_{\rm H} \frac{\dust}{\dsun}\,\cc\,{\rm s}^{-1}\,,
\ee
where $n$ and $n_{\rm H}$ are the total and hydrogen gas densities, respectively. Note that the dust-channel is dominant with respect to the gas-phase formation channel for $\dust\gsim 10^{-2}\dsun$, i.e. for most of the lifetime of a typical EoR galaxy \citep{pallottini:2017_b}.

Cosmic-rays are not explicitly tracked in the simulation \citep[cfr. ][]{dubois:2016,pfrommer:2017}. Similarly to what is done in \citet{pallottini:2017_b,pallottini:2019}, we assume a cosmic-ray hydrogen ionisation -- and associated Columb heating -- rate proportional to the global SFR \citep{valle:2002} and normalised to the MW value \citep[][see \citealt{ivlev:2015, Padovani18} for the spectral dependence]{webber:1998}, i.e. $\zeta_{\rm cr} = 3\times 10^{-17} ({\rm SFR}/\msunyr)\, {\rm s}^{-1}$. Note that the global SFR is computed by averaging over the stellar mass formed in the whole simulation in the last $20\,\myr$, and it is typically dominated by the target galaxy in the zoom-in region \citep[see also][]{pallottini:2017}.
Note that this likely overestimates the ionization and heating from cosmic-rays; while it is possible to track the cosmic-ray field consistently \citep{farcy:2022}, we leave such modification for a future work.

Using on-the-fly radiative transfer, absorption is done on a cell by cell basis. The process depends on the abundance of each single species, dust amount, and the selected MW grain size distribution\footnote{A different assumption on the dust distribution can give relevant effects on the observed SED \citep{behrens:2018}, however it typically gives only minor differences for the chemical network; for instance adopting a Small Magellanic Cloud or MW composition have average differences of $\simeq 1\%$, that goes up to $\simeq 30\%$ for energy bins that contribute neither to \HH~dissociation nor to H~ionization \citep{pallottini:2019}.}. Cross-sections for gas processes are directly taken from the reaction rates of the network. For photoreactions, in each radiation bin the relative cross-section is pre-computed and averaged on the SED of the fiducial stellar population. The self-shielding of \HH~from photo-dissociation is accounted given the \HH~column density, temperature and turbulence \citep{richings:2014}.

\code{krome} solves the chemical and thermal evolution by adopting \code{DLSODES} \citep{hindmarsh:1983}, an implicit solver that takes advantage of the sparsity of Jacobian matrix associated with an astrophysical chemical network. As noted in \citet{pallottini:2019}, in its current version \code{krome} evolves chemical species by assuming that the radiation field is constant in a time step; to account for absorption associated with chemical processes, we sub-cycle the absorption step in order to assure convergence \citep[][in particular see the benchmarks in their Appendix]{decataldo:2019}.

\subsubsection{Gas thermodynamics}

We model both the evolution of thermal and turbulent (or non-thermal) energy content of the gas.

Thermal energy can change due to radiative cooling and heating which are computed using \code{krome} and accounting for the local ISRF. Extra thermal energy inputs are due to stellar feedback (e.g. SNe and winds, see Sec. \ref{sec_stellar_feedback}).
Since metal species are not followed individually, we use the equilibrium metal line cooling function calculated via \code{cloudy} \citep{ferland:2013} with a \citet{haardt:2012} UV background\footnote{This is inconsistent, for a discussion see \citet{pallottini:2017_b} and \citet[][in particular App. D]{emerick:2019}, the latter showing that the largest difference can be found for gas with $0.1 \lsim n/\cc \lsim 1$.}. Following cooling from individual metal species can modify the thermodynamics of the low density ISM \citep[][]{gnedin:2012,capelo:2018}, since the cooling function typically changes by a factor $\lsim 2$ \citep{bovino:2016}. Such effect might be important to correctly compute emission lines strengths, particularly for some lines \citep{lupi:2020}. To overcome this problem and to reach a better accuracy, in \code{serra} we perform this computation in post-processing (Sec. \ref{sec_emission_line} and \ref{sec_emission_continuum}).
Dust cooling is not explicitly included since it gives a minor contribution to the gas temperature for $n<10^4\,\cc$ \citep[][see their Fig.~3]{bovino:2016}.
Note that the cosmic microwave background (CMB) effectively sets a temperature floor for the gas. Its inclusion is fully described in \citet{pallottini:2017_b}.

The non-thermal energy density $e_{\rm nth}$ of the gas is treated as a passively advected variable \citep{agertz:2015}, that is sourced by stellar feedback (e.g. SNe, winds, and radiation pressure, see Sec. \ref{sec_stellar_feedback}). Once turbulence is injected in the gas, it can dissipate at a rate \citep[][see eq. 2]{teyssier:2013}
\be
 \dot{e}_{\rm nth} = - \frac{e_{\rm nth}}{ t_{\rm diss}}\,,
\ee
where $t_{\rm diss}$ is the dissipation time scale. This can be written as the eddy turn-over scale \citep{mac_low:1999}
\be
 t_{\rm diss} = 9.785 \left( \frac{\Delta l}{100\,{\rm pc}}\right)\left(\frac{\sigma_{\rm t}}{10\,{\rm km}\,{\rm s}^{-1}}\right) ^{-1}\,\myr\,,
\ee
where $\sigma_{\rm t} = \sqrt{e_{\rm nth}}$ is the turbulent velocity dispersion.
More refined turbulence models are available \citep{scannapieco:2010,iapichino:2017,engels:2019}, and have been successfully used in zoom-in simulation \citep{semenov:2016,semenov:2018}. However, they have not been adopted in the present since the resulting SFR efficiencies are comparable \citep[][]{pallottini:2017}.

\subsection{Stars}\label{sec_stars}

Stellar particles in the simulation have a minimum mass of $m^{\rm min}_\star = 1.2\times 10^4 \msun$, the gas mass resolution in the zoom-in region. Thus, a single star particle in our simulations should be rather seen as a proxy for a \quotes{stellar cluster}. We assume a $0.1-100\,\msun$ \citet{kroupa:2001} Initial Mass Function (IMF). We keep track of the metallicity and age of each stellar particle. This information is used to compute chemical, radiative and mechanical feedback via \code{starburst99} \citep{leitherer:1999}, by adopting evolutionary tracks of stellar population from \code{padova} \citep{bertelli:1994} library, that covers the $0.02 \leq Z_{\star}/\zsun \leq 1$ metallicity range.

\subsubsection{Star formation}\label{sec_formation}

Stars form according to a Schmidt-Kennicutt relation \citep[][]{schmidt:1959,kennicutt:1998} that depends on the molecular hydrogen density ($\nHH$):
\be\label{eq_sk_relation}
\rho_{\rm SFR}= \zeta_{\rm SFR} \frac{\mu m_{\rm p} \nHH}{t_{\rm ff}},
\ee
where $\rho_{\rm SFR}$ is the local star formation rate density, $\zeta_{\rm SFR}$ the efficiency, $m_{\rm p}$ the proton mass, $\mu$ the mean molecular weight, and $t_{\rm ff}$ the local free-fall time.
The efficiency is set to $\zeta_{\rm SFR}=10\%$, i.e. adopting the average value observed for MCs \citep[][see also \citealt{agertz:2013}]{murray:2011}.

Eq. \ref{eq_sk_relation} is solved stochastically at each time step $\Delta t$ in each cell (size $\Delta l$) \citep{rasera:2006,dubois:2008}, by forming a new star particle with mass $m_\star = m^{\rm min}_\star\,N_\star $, with $N_\star$ an integer drawn from a Poisson distribution characterised by mean
\be\label{eq_mean_poisson_sfr}
\langle N_\star\rangle = \zeta_{\rm SFR} \frac{\mu m_{\rm p} \nHH \Delta l^3}{m^{\rm min}_\star} \frac{\Delta t}{t_{\rm ff} }\,.
\ee
For numerical stability no more than half of the cell mass is allowed to turn into stars. The stellar metallicity, $Z_{\star}$, is set equal to the gas metallicity, $Z$, of the spawning gas cell.

\subsubsection{Stellar feedback}\label{sec_stellar_feedback}

Stellar energy inputs, chemical yields, and photon production depend both on metallicity $Z_\star$ and age $t_\star$ of the stellar cluster, according to the \code{padova} tracks. Stellar feedback includes SNe, winds from massive stars, and radiation pressure \citep[see also][]{agertz:2013}.

Depending on the kind of feedback, stellar energy input can be both thermal and kinetic, and we account for the dissipation of energy in MCs for SN blastwaves \citep{ostriker:1988} and OB/AGB stellar winds \citep{weaver:1977}. The relative fraction of thermal and kinetic energy depends on the SN blast stage: energy conserving Sedov-Taylor stage (about 70\% thermal, 30\% kinetic), shell formation stage, and pressure driven snowplow (about 15\% thermal and 35\% kinetic). This is detailed in Sec.~2.4 and Appendix A of \citet{pallottini:2017}.

Note that radiation pressure is implemented by adding a source term to the turbulent (non-thermal) energy, as detailed in \citet{pallottini:2017}. Thus, to avoid double counting of such feedback, we turn off the standard radiation pressure prescription of \code{ramses-rt} \citep[][]{rosdahl:2015}.

\subsection{Post-processing}\label{sec_finders_emission}

After each individual zoom-in simulation is completed, we identify galaxies and compute their emission properties. Each simulation outputs multiple snapshots equispaced in redshift ($\Delta z\simeq 0.5$) for $15 \leq z \leq 9$ and equispaced in time ($\Delta t \simeq 10 \,\myr$) for $9 \leq z \leq 6$; this setup gives a total of about 70 snapshots per simulation. In each snapshot, apart from the central target, there are about $10-30$ galaxies in the zoom-in region, depending on environment and redshift.

\subsubsection{Identification}\label{sec_identification}

To identify DM halos and galaxies we use \code{rockstar-galaxies}\footnote{\url{https://bitbucket.org/pbehroozi/rockstar-galaxies}} \citep{behroozi:2013}, a clustering algorithm that performs a phase-space identification for multi-mass simulations. It uses a friend-of-friend algorithm that adopts a metric accounting for both spatial and velocity separations.
We treat separately DM and stellar particles, and allow \code{rockstar-galaxies} to identify groups with a minimum number of $100$ ($20$) DM (stars) particles. While \code{rockstar-galaxies} provides the linking relations between sub-halo (satellites) and halo (main galaxy), in the present paper we treat each galaxy as an individual object. We note that using a typical parameterisation for the phase-space distances is too coarse to allow a complete and robust identification of dwarf satellites of high-$z$ galaxies \citep[cfr.][]{gelli:2020}.

To identify progenitors/descendants through cosmic time, we use an algorithm similar to \code{mergertree} \citep{knebe:2013}. First, we produce all the galaxy/DM halo catalogues of each snapshot. For each consecutive couple of snapshots, we cross match the galaxies/DM halos identified in the two catalogues.

Given two snapshots at redshift $z_1 < z_2$, the halo A at $z_1$ is flagged as a progenitor of the halo B at $z_2$ if their intersection has $N_{A \cap B} > 100$ DM (20 stellar) particles. The main progenitor of B is the halo A that maximises the merit function $\mathcal{M}_1 = N^2_{A \cap B}/N_{A} N_{B}$ \citep{chaichalit:2013}. Using the progenitor list for each couple of consecutive snapshots, we can build the full progenitor/descendant graph for each galaxy and halo.
If two galaxies keep exchanging mass (e.g. during a merger spanning multiple snapshots), we supplement our progenitors search by cross-matching the progenitor catalogue with the descendant catalogue. The latter is obtained by looking for halos in snapshot $z_2$ that are descendant halos in snapshot $z_1$, with the same procedure outlined above.
Considering only particle IDs in the progenitor/descendant identification is prone to a potential misidentification of progenitors/descendants \citep{chaichalit:2013} and can be avoided by adding dynamical information (see e.g. \code{consistent-trees} from \citealt{behroozi:2013_b}). However, i) the fine time sampling of our snapshots ameliorate the problem and ii) we scarcely use such information in the present study, thus we will consider such treatment in future work.

Finally, we label a galaxy by the name of the simulation, number of the snapshot, and stellar halo ID, and we often indicate it with a specific name.\footnote{When a specific name is chosen for a galaxy, it is propagated through the main progenitor and descendant branch of its graph. For instance, \quotes{Freesia} is the most massive galaxy in \texttt{serra00}: it has been identified at $z\simeq 8$ and its main characteristics have been described in \citet{pallottini:2019}}.

\subsubsection{Line emission}\label{sec_emission_line}

Line emission is calculated on a cell by cell basis by using \code{cloudy} models. We use grids of \code{cloudy} models for density, metallicity, radiation field intensity, as a function of the column density, which is used as a stopping criterion for the calculation. We account for the turbulent and clumpy structure of the interstellar medium (ISM), by parameterising the underlying distribution as a function of the gas Mach number \citep{vallini:2017,vallini:2018}, expressing the thermal to turbulent energy ratio.
We have two grids of \code{CLOUDY} models, in which the impinging SED includes -- does not include -- ionising radiation ($h\nu > 13.6$ eV). We consider radiation to be ionising in those cells that i) contain young ($t_\star\leq 10\,\myr$) stars or ii) have a ionisation parameter larger then $\geq 10^{-3}$.

Every grid is composed of 17 density bins ($10^{-2} \le n/\cc \le 10^{6.5}$), 8 metallicity bins ($10^{-3} \le Z/\zsun \le 10^{0.5}$), and 12 ISRF intensity bins ($10^{-1} \le G/\gnot \le 10^{4.5}$), for a total of 1632 distinct models per grid.
For the SED of the impinging radiation field on the slab of gas of interest in \code{CLOUDY}, we use our fiducial SED with stellar age of $10\,\myr$ and solar metallicity, i.e. the stellar population is the primary contributor to the interstellar radiation field in our simulated galaxies. Note that in the post-processing via \code{cloudy}, the spectrum is not limited to 24.59 eV. The full setup is described in \citet{pallottini:2019}, while in \citet{lupi:2020} a detailed analysis of different emission line models is reported \citep[see][for additional discussion]{olsen:2018}.

Line emission products can be spatially resolved maps or Hyperspectral Data cubes, i.e. line spectra in a position -- position -- line of sight velocity space \citep{kohandel:2020}. Given a field of view and a line of sight, for optically thin lines (e.g. FIR lines such as \CII~158$\mu $m) we can directly sum the contribution of all the gas cells in the field of view. For optically thick lines (e.g. \CIII~1909$\angstrom$) we should consider the radiative transfer trough dust \citep[see][for the additional efforts needed for Ly$\alpha$]{behrens:2019}.
Unattenuated (de-reddened) emission lines can be readily calculated and approximate treatments for the attenuation can give reliable estimates \citep{gelli:2021}, however we defer more refined implementations to future work.

\subsubsection{Continuum emission}\label{sec_emission_continuum}

Continuum emission is generated by using \code{skirt}, a Monte Carlo based code that computes the radiative transfer process in dusty media. The setup adopted here is fully described in \citet{behrens:2018} and we summarise it as follows.

The spatial distribution of the light sources is taken from the position of the stellar clusters (particles) in the simulations; for each cluster, we use its metallicity and age to compute the stellar SED, by adopting \citet{bruzual:2003} models and a \citet{kroupa:2001} IMF. We adopt $10^7$ photon packages per wavelength bin per source, which ensure a good convergence \citep[see][, in particular App. A]{behrens:2018}.
While this is taken as our fiducial model, we note that using different stellar emission models and/or including the nebular continuum can change intrinsic UV broadband properties by a factor up to $\simeq 20\%$ \citep{wilkins:2016_b}.

Dust distribution is taken from the simulations by adopting a fiducial value of $f_\mathrm{d}=0.08$ \citep{behrens:2018} for the dust-to-metal ratio, i.e. the best-fit simultaneously matching the UV and FIR SED of selected high-$z$ galaxies \citep{laporte:2017}. We also assume a dust composition and grain size distribution reproducing the MW extinction curve \citep{weingartner:2001}, and a dust emissivity $\beta_d = 2$. The value of $f_d$ and dust composition are uncertain at high-$z$ \citep{wiseman:2017}. Note that in the \code{skirt} runs, we are adopting a $f_d$ value that is $3.75$ times higher than the one used in the simulations. In the future we reserve the option to change such fiducial setup, e.g. from $f_\mathrm{d}=0.08$ to $f_\mathrm{d} \simeq 0.2$ as seen in local galaxies \citep{delooze:2020}, or use the MW $f_\mathrm{d} = 0.3$ value \citep[e.g.][]{hirashita:2002}.
We use 100 logarithmically spaced spectral bins to cover the $-1\leq \log(\lambda/\mu m) \leq 3$ restframe wavelength and an additional finer grid of 40 bins at $0\leq \log(\lambda/\mu m) \leq 1.5$, i.e. to have a good coverage of the MIR range.
Using \code{skirt} we compute the scattering and absorption without accounting for dust self-absorption, which can be relevant for column densities of the order of $\gsim 10^{24}{\rm cm}^{-2}\dsun/\mathcal{D}$, but gives only a minor contribution to the mid-infrared emission in our typical object \citep[][in particular see App. A]{behrens:2018}.

Similarly to the emission lines, the continuum data products are spatially resolved spectra.

\section{Results}\label{sec_results}

\begin{figure*}
\centering
\includegraphics[width=0.95\textwidth]{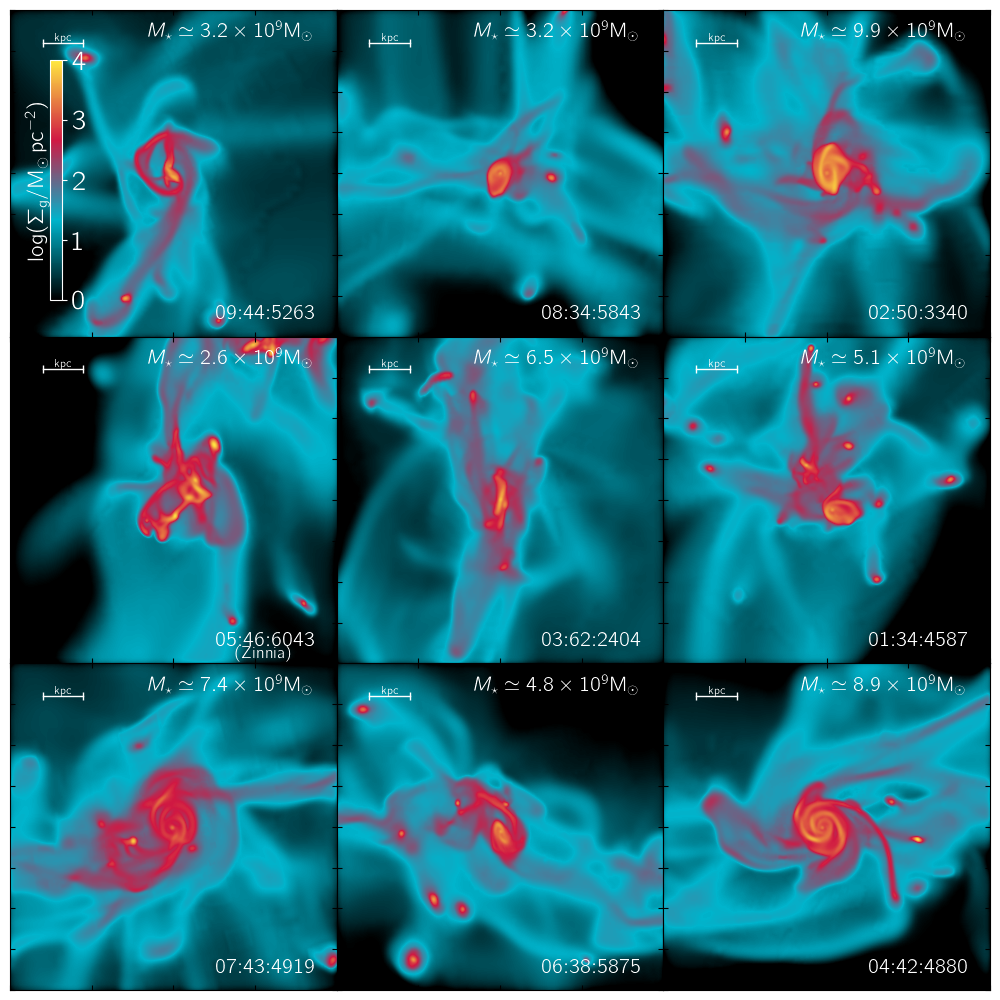}
\caption{
Overview of \code{serra} galaxies.
Each panel shows the gas surface density ($\Sigma_g$) of a galaxy with $M_\star \sim 10^{9}\msun$. To give a qualitative overview of the different morphologies found in \code{SERRA}, galaxies are taken at $6\lsim z\lsim 8$, each coming from a different simulation in the suite. For each panel, the galaxy stellar mass and identifier (see Sec. \ref{sec_identification}) and name when present are shown in the upper right and lower right parts, respectively.
\label{fig:warhol_overview}}
\end{figure*}

What do \code{serra} galaxies look like? To have a visual representation, we select 9 different $M_\star \sim 10^9\msun$ galaxies in the redshift interval $6\lsim z \lsim 8$, chosen as a showcase of the different evolutionary stages. In Fig. \ref{fig:warhol_overview} we plot maps of the gas surface density ($\Sigma_g$) of these galaxies, oriented face-on\footnote{We define the face-on orientation by using the total stellar angular momentum. This working definition gives reasonable results even in the case of a close merger.}.

In a cosmological structure formation scenario, gas starts to condense in overdensities generated by dark matter halos; via filamentary accretion in the IGM/CGM, gas flows to the centre of the potential DM well, where it can cool and reach densities high enough that \HH~and thus star formation is possible (upper panels in Fig. \ref{fig:warhol_overview}).
Mergers represent landmarks of galaxy evolution, as per the hierarchical build-up scenario, and objects can often be seen in different phases of the merging process (middle panels in Fig. \ref{fig:warhol_overview}).
If the galaxy mass is high enough and it is not disturbed by mergers, a spiral structure \citep{ceverino:2015,pallottini:2017} forms within a rotating and pressure supported disk \citep{inoue:2016,leung:2020}, while it continues to accreting gas and forming stars (lower panels in Fig. \ref{fig:warhol_overview}).

The morphological evolution is particularly rapid at high-$z$, as the rate of mergers per unit time and fresh gas accretion is an increasing function of redshift \citep{fakhouri:2010,correa:2015}. 
Furthermore, the stellar feedback is particularly violent because of the episodic and bursty nature of the star formation history \citep{pallottini:2017_b,ma:2018}. Thus the same galaxy can dramatically change its morphology in a relatively short time lapse, i.e. $\sim 25\,\myr$ \citep[e.g.][Fig. 5]{kohandel:2019}.

\subsection{Stellar build-up}\label{sec_analysis_evolution}

\begin{figure}
\centering
\includegraphics[width=0.49\textwidth]{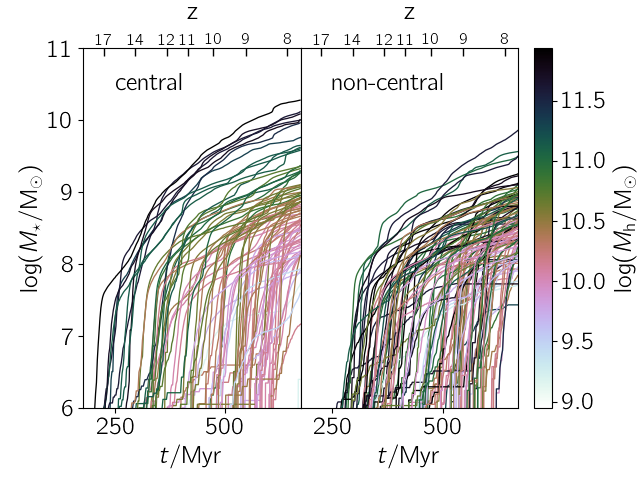}
\caption{Cumulative stellar mass ($M_\star$) as a function of cosmic time ($t$) in the \code{serra} simulation. Galaxies are divided into central (89 objects, left panel) and non-central (113 objects, right panel), depending on whether they are the most massive galaxy in the host dark matter halo.
For each of the 202 galaxies, the evolutionary track is coloured (see the colorbar) according to the host halo mass\footnoteref{def_m200} at $z=7.7$.
The redshift ($z$) corresponding to $t$ is indicated in the upper axis.
\label{fig:time_evolution_halo}}
\end{figure}

\begin{figure*}
\centering
\includegraphics[width=0.49\textwidth]{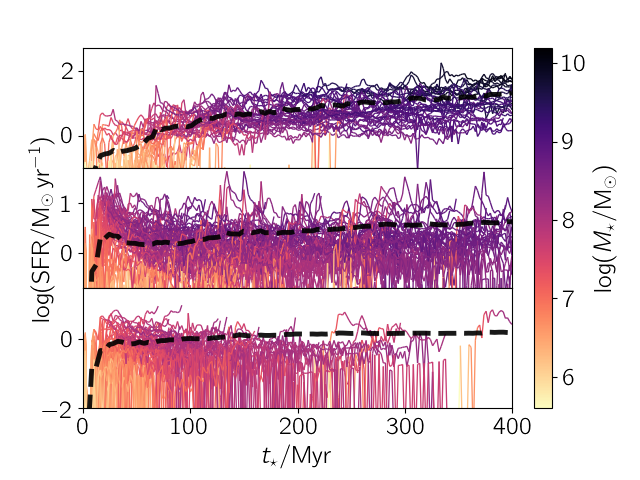}
\includegraphics[width=0.49\textwidth]{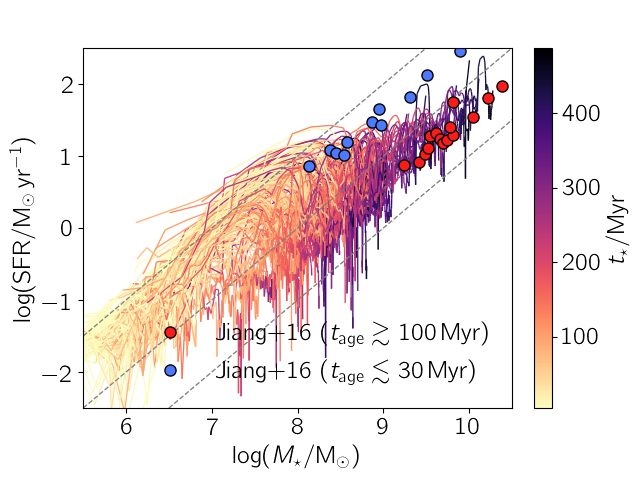}
\caption{
Star formation history in \code{serra}.
\textbf{Left panel}: star formation rate (SFR) as a function of stellar age\footnoteref{def_age} ($t_\star$). Each galaxy track is coloured with $M_\star$ at $z=7.7$. The data is divided into 3 samples: a galaxy is in the upper panels if $\SFR > 13.8 \,\msunyr$ at least once during its history, in the middle panel if $\SFR \gsim 4.8 \,\msunyr$, in the lower panel otherwise. Limits are chosen such that each panel contains about 67 galaxies. The SFR average for each subsample is shown by the black dashed line in each panel. Individual $\SFR$ (averages for each subsample) are computed over a period of $2\,\myr$ ($10\,\myr$).
\textbf{Right panel}: evolution of $\SFR$ vs $M_\star$. Each galaxy is coloured with its age as it increases with $M_\star$. Crosses are the observations from \citet{jiang:2013} with the sample divided in young and old galaxies. The three dashed lines indicate a constant value of $\sSFR \equiv \SFR/M_\star = 1,\,\,10,\,100\, \invgyr$. The $\sSFR$ has been averaged over $20\,\myr$ to reduce flickering effects.
\label{fig:time_evolution_galaxy}}
\end{figure*}

We start the quantitative analysis of \code{serra} by looking at the formation history of the stellar component of our galaxies.

The stellar mass build-up of galaxies in \code{serra} is plotted in Fig. \ref{fig:time_evolution_halo}.
The evolution is shown up to $z=7.7$ for 202 individual galaxies, each one of them being coloured according to the host halo mass\footnote{Halo mass definition can be done either by direct summation of particle groups in the halo finder or using a spherical overdensity criterion. Here as halo mass we adopt \quotes{M200c}, i.e. integrating the DM mass contained in a sphere centred around the stellar centre of mass up to the radius where the density is 200 times higher then critical density $\delta_c$, i.e. $M_{\rm h} = \int \rho_{dm} {\rm d}^3 r = 200 \delta_c \int {\rm d}^3 r$. Selecting a different threshold values for the density (e.g. $\Delta_c$ from the spherical collapse model or 500) or adopting the matter density instead of the critical density can give differences up to a factor 3 in the halo mass.\label{def_m200}}. Note that all the galaxies in Fig. \ref{fig:time_evolution_halo} are defined via object identification at $z=7.7$, thus mass contribution from $z>7.7$ mergers is included in the evolutionary track.

Galaxies start to form from $z\simeq 18$, when the Universe is $t \simeq 200$ Myr old. In the plotted redshift range the stellar masses span from $10^{6}$ to $5\times 10^{10} \msun$ and the galaxies have typical metallicity in the range $0.1 \lsim Z/\zsun \lsim 0.5$ \citep[see also][]{gelli:2022}..
Since the suite is surveying biased regions of the Universe, relatively rare galaxies ($M_\star \gsim 10^{10} \msun$) are present since $z\lsim 9$; in particular, 37 targets have $M_\star \geq 10^{9} \msun$ at $z=7.7$.

Objects in our sample can be divided into central (left panel of Fig. \ref{fig:time_evolution_halo}) and non-central (right panel), depending on if they are the most massive galaxy in the hosting DM halos. For central galaxies, the higher the DM halo mass, the higher are the formation redshift of the first stellar populations, as a consequence of the hierarchical structure formation process.

For non-central galaxy, a relation between $M_\star$ and $M_h$ is not expected; note that non-central galaxies are not necessarily satellites of the most massive galaxy in the hosting DM halo; however, being at a relatively close distance ($\lsim 20\,\kpc$), they can be affected by mass exchange, metal enrichment, and radiative feedback from the latter. In the following Sec.s we see that they typically follow similar scaling relations as the central galaxies but with a larger dispersion.

Regardless of the distinction, almost all the galaxies experience a rapid increase of their stellar mass in their first $\sim 100\,\rm Myr$, followed by a less intense growth phase. To have a closer look, we can check their star formation history (Fig. \ref{fig:time_evolution_galaxy}).
In the left panel, we plot the star formation rate (SFR) as a function of galaxy age\footnote{The galaxy age $t_\star$ is defined as the time from the first star formation event; we stress it should be interpreted as time reference more then a galaxy physical property \citep[cfr][]{tacchella:2018}.\label{def_age}} ($t_\star$).
Data for both central and non-central is divided into 3 groups of 69 targets, depending on the maximum SFR reached during the history of the galaxy, and each galaxy track is coloured with the stellar mass at $z=7.7$.

During the first $400 \,\myr$, galaxies with $M_\star \lsim 10^7 \msun$ typically have $\SFR \lsim 5\, \msunyr$, intermediate mass galaxies ($10^8 \lsim M_\star \lsim 10^9 \msun$) wiggle around $\SFR \simeq 10\, \msunyr$, while $M_\star \gsim 10^9 \msun$ have $\SFR \simeq 30\, \msunyr$ for long spans of times, and can reach peaks of $\SFR \sim 200\, \msunyr$ during short bursts.
The majority of bursts during the galaxy history have time scales of the order of $\simeq 10\, \myr$: a peak in SFR activity determines a delayed peak of stellar feedback, that temporarily suppresses the star formation \citep[see][for more details regarding the feedback impact]{pallottini:2017}. Such suppression typically becomes a temporary quenching in low mass galaxies \citep{gelli:2020}. More details on this effect are given later on. Note that extreme variations -- bursts of the order of $10\times$ the time averaged value -- are likely induced by merger events and massive inflow of new fresh gas.

\subsection{Main sequence}\label{sec_main_sequence}

It is interesting to look at the history in the SFR$-M_\star$ plane, i.e. the \quotes{main sequence}, that is shown in the right panel of Fig. \ref{fig:time_evolution_galaxy}.
Galaxies in \code{serra} start their life with a specific star formation rate $\sSFR \equiv \SFR/M_\star\simeq 100 \,\invgyr$ and gradually move to $\sSFR\simeq 10 \,\invgyr$ as they grow older \citep[around $t_\star \simeq 200\,\myr$, see also][]{pallottini:2017_b}.
As expected from the sSFR evolution \citep[see][and reference within]{madau:2014}, galaxies in the EoR have extreme values with respect to what observed locally ($z=0$, $\sSFR\simeq 0.1 \,\invgyr$) and at low redshift ($z\simeq 1$, $\sSFR\simeq 1 \,\invgyr$).
The values reported for $M_\star \simeq 10^9\msun$ galaxies in \code{serra} are in good agreement with those observed at $z\sim 7$ \citep{gonzalez:2010,stark:2013}. Moreover, $z\simeq 6$ observations by \citet{jiang:2013} show that young (old) galaxies have ${\rm sSFR} = {\rm SFR}/M_{\star} = 39.7\, \gyr^{-1} (4.1\, \gyr^{-1})$, i.e. hinting at a decreasing trend of sSFR with galaxy age.
Direct comparison with the data from \citet{jiang:2013}, reveals that the general trends from the stellar tracks of \code{serra} galaxies well reproduce the observations of old galaxies. However we find ${\rm sSFR} = 39.7\, \gyr^{-1}$ for young $M_\star \simeq 10^9 \msun$ galaxies only during bursty periods.
Bi-modality in the $M_\star$-SFR plane has been probed by observations of large samples up to $z\lsim 6$ by \citet{rinaldi:2021}, however the author show that \code{IllustrisTNG50} simulations \citep[][]{pillepich:2019} struggle to reproduce the starburst cloud at any redshift interval.
Further and more refined analysis will be needed in order to quantify if this lack is due to intrinsic limitations in the simulations or to a bias in the comparison (e.g. because of the volume).

\subsection{Halo-stellar mass relation}\label{sec_analysis_z_7_7}

\begin{figure}
\centering
\includegraphics[width=0.49\textwidth]{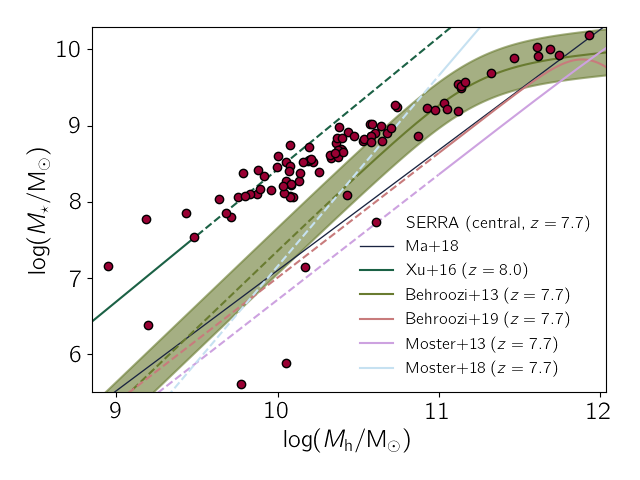}
\caption{
Halo - stellar mass ($M_h-M_\star$) relation for 89 simulated central galaxies at $z=7.7$. We overplot the results from other numerical simulations from \code{Renaissance} \citep[][at $z=8$]{xu:2016} and \code{FIRE-2} \citep[][redshift independent fit]{ma:2018}, along with the results from different abundance matching models \citep[][]{behroozi:2013_c,behroozi:2019,moster:2013,moster:2018} at the same redshift. Dashed lines indicate the $M_h$ ranges where models are extrapolated. In this work, \citet{behroozi:2013_c} is the reference for comparisons, the shaded region shows the standard deviation.
\label{fig:halo_mass_stellar_mass}
}
\end{figure}

\begin{figure}
\centering
\includegraphics[width=0.49\textwidth]{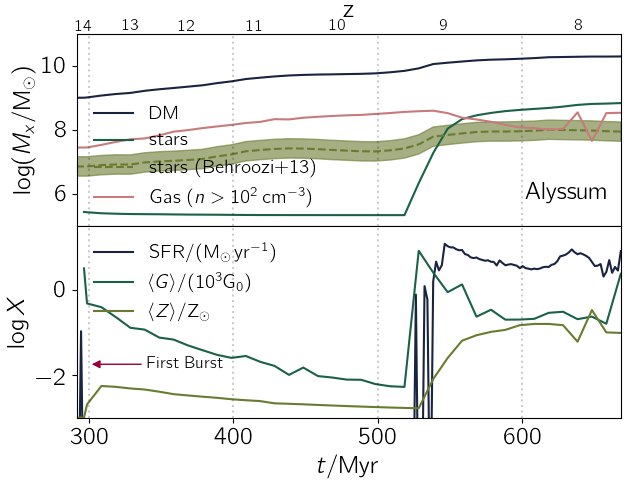}
\caption{Evolutionary history of \quotes{Alyssum}, a $M_\star \simeq 5\times 10^8\msun$ \code{serra} galaxy at $z=7.7$.
{\bf Upper panel}: Dark matter (DM), stellar, and dense gas ($n>10^2\cc$) mass as a function of cosmic time and corresponding redshift; for reference, we add the evolution of $M_\star(M_h,z)$ from \citet{behroozi:2013_c}.
{\bf Lower panel}: Star formation rate (SFR), average Habing radiation field ($G$), and gas metallicity ($Z$).
Note that the both SFR and $M_\star$ are computed with a $\Delta t = 2 \, \rm Myr$ time sampling, while the other quantities have a coarser $\Delta t \simeq 10 \, \rm Myr$ sampling, due to availability of simulation snapshots.
\label{fig:alyssum_hystory}}
\end{figure}

Here we focus our analysis at $z=7.7$ and we start by looking at the halo mass to stellar mass relation, that is often used as a benchmark for the efficiency of star formation.
The relation for central galaxies in \code{serra} is presented in Fig. \ref{fig:halo_mass_stellar_mass}.
At $z=7.7$ the halo masses range from about $10^{9}$ to $5\times 10^{11}\msun$ and the stellar mass of central galaxies from about $10^6$ to $5\times 10^{10}\msun$.
Few objects with mass of $\simeq 10^6 \msun$ are present, as most of them are embedded in halos hosting more massive galaxies, while smaller objects are not identified as galaxies, because of the threshold of 20 stellar particles adopted in the definition (see Sec. \ref{sec_identification}).

In the same figure we compare with the results from abundance matching models \citep{behroozi:2013_c,behroozi:2019,moster:2013,moster:2018} and the $M_\star-M_{h}$ from numerical simulations \citep{xu:2016,ma:2018}.

The central galaxies in \code{serra} have an average trend that is in broad agreement with the $M_\star-M_{h}$ from \citet{behroozi:2013_c}. Notably, the redshift-dependent turnover at $M_h\sim 10^{11}\msun$, corresponding to the turnover at $L_\star$ in the luminosity function, is well reproduced. The \code{serra} predictions diverge from the \citet{behroozi:2013_c} curve at $M_h \lsim 5\times 10^{10}\msun$: most of the galaxies show a larger stellar mass, with few blatant outliers that sit well below it.
Qualitatively a similar behaviour is found when comparing with \citet{moster:2013}, although at $z=7.7$ there is a disagreement at more than $1\sigma$ around $M_h\sim 10^{11}\msun$ between \citet{moster:2013} and \citet{behroozi:2013_c}; such a disagreement is mainly due to the different assumption in the abundance matching models and the different extrapolation methods \citep[see also][]{behroozi:2015,moster:2018}.
In particular, for $M_h \lsim 10^{10} \msun$ and at $z\gsim 6$ \citet{behroozi:2013_c} should be taken with care, since results in that range have been extrapolated due to the lack of data. The situation is qualitatively similar also by considering updated versions of the abundance matching models \citep{moster:2018,behroozi:2019}.

Despite these caveats, in the present work we take the $M_\star-M_{h}$ from \citet{behroozi:2013_c} as a reference. With this considerations in mind, it seems that feedback at the low mass end in \code{SERRA} is too weak to suppress star formation significantly and/or the lack of resolution to solve the mini-halo progenitors is spoiling the evolution of low mass galaxies.

As \citet{behroozi:2013_c} is broadly consistent with the relation from \citet{ma:2018}, similar considerations hold for the comparison between \code{serra} and \code{FIRE-2}. However, note that the latter reports no redshift evolution for the $M_\star-M_{h}$, while -- particularly at high redshift -- there is a rapid evolution in the $M_\star-M_{h}$ from \citet{behroozi:2013_c}.
In the low-mass end results from \code{SERRA} are closer to what reported by \code{Renaissance} as analysed in \citet{xu:2016}; note that we report the fit from \citet{xu:2016} extrapolating at $M_h > 10^{10} \msun$, so extra care should be taken in looking at the high mass end from \code{Renaissance}.

To summarise, while at high $M_h$ the \code{serra} results are consistent with different models, at low $M_h$ they generally favour larger stellar masses with respect to the extrapolated predictions. How can we understand such a trend?

\subsubsection{Low-mass galaxies: interpreting their trend}\label{Sec:Alyssum}

To answer this question, consider the star formation history (Fig. \ref{fig:alyssum_hystory}) of \quotes{Alyssum}, a typical central, low-mass ($M_\star \simeq 5\times 10^8\msun$) \code{serra} galaxy\footnote{Note that for the analysis of Alyssum we relax the threshold of a minimum 20 particles to define a galaxy: for snapshot prior to the first identification of the galaxy with \code{rockstar}, we backtrack each single star particle to the first formation event.} at $z=7.7$. Alyssum does not experience any major merger down to $z=7.7$. Its first star formation event occurs in an almost pristine ISM at $z\simeq 14$, when the galaxy halo mass is $M_h\simeq 10^9\msun$ or $T_{\rm vir} \simeq 10^4$ K. 

In the first burst, a $M_\star\simeq 5\times 10^5 \msun$ stellar cluster is formed; the star formation is completely quenched for a time span of 200 Myr thereafter. The metallicity rapidly rises from the initial floor value of $Z= 10^{-3}\zsun$ to $Z\simeq 10^{-2}\zsun$, and subsequently slowly decreases because of the onset of a powerful outflow, and -- most prominently -- cosmological accretion of fresh gas.

Quenching of star formation is due to radiative feedback. Initially, the mean intensity of the \HH\, photo-dissociating radiation field\footnote{$G_0 = 1$ roughly corresponds to $J_{21}=100$, where $J_{21}$ is the radiation intensity in units of $10^{-21} \rm erg\, s^{-1}\, cm^{-2}\, Hz^{-1}\, sr^{-1}$.} produced by the newly born stars is very high, $\langle G \rangle=10^3 G_0$ (see lower panel of Fig. \ref{fig:alyssum_hystory}). This is well above the critical threshold for \HH\, formation in a pristine environment, $\langle G^* \rangle\simeq 10\, G_{0}$ \citep{johnson:2013_b}. Thus, the galaxy cannot form stars until the ISRF decreases below this threshold, roughly 200 Myr after the burst. In this first phase, $M_\star$ is below the (extrapolated) \citet{behroozi:2013_c} relation.

When star formation is reignited, the process repeats with increasingly shorter cadence. This is due to the fact that metals and dust produced by the SNe improve the self-shielding ability of the gas, thereby enabling \HH\, formation even in the presence of a super-critical ISRF intensity. After two additional bursts, the star formation stabilises to $\simeq 10 M_\odot {\rm yr}^{-1}$, and steadily continues in spite of the very strong, $\langle G\rangle\simeq 100\, G_{0}$, ISRF. The metallicity has overtaken the $Z = 10^{-2} \zsun$ level, which we can therefore empirically identify as the minimum metallicity required to sustain a continuous star formation activity. At that point the galaxy shortly overshoots the \citet{behroozi:2013_c} $M_\star-M_h$ curve. This situation is however transient: more massive galaxies later on fall back on the $M_\star-M_h$ (Fig. \ref{fig:halo_mass_stellar_mass}).

The described trend crucially depends on the assumption that star formation occurs in molecular clouds. We cannot exclude that a different star formation mode, occurring in dense -- but not necessarily molecular -- gas \citep[e.g.][]{semenov:2016,lupi:2018} might be at work in early galaxies. In such case, the quenching phase could be shorter or even completely absent. The quenching details and duration might also depend on mass resolution, although the physical arguments derived from the simulation analysis should hold anyway. We will explore these possibilities in future work.

\subsection{Schmidt-Kennicutt relation}\label{sec_sk_relation}

\begin{figure}
\centering
\includegraphics[width=0.49\textwidth]{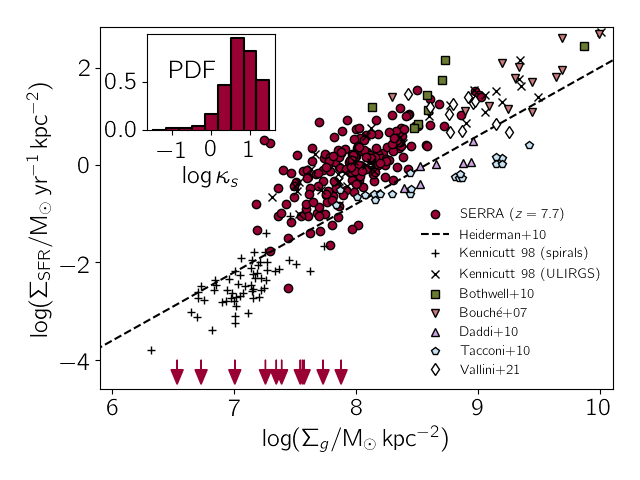}
\caption{
Schmidt-Kennicutt relation at $z=7.7$. We plot the \quotes{classical} relation \citep[][]{krumholz:2015}, i.e. star formation, $\sigmasfr$, vs. gas, $\Sigma_g$, surface density.
The data from \code{serra} is marginally resolved, i.e. we smoothed the maps of both $\sigmasfr$ and $\Sigma_g$ with a Gaussian kernel with a width of $0.5$ kpc and take the maximum/central values in the maps, in order to have a uniform sample.
For the 11 objects that are temporarily quenched at $z=7.7$ (see Sec. \ref{sec_analysis_z_7_7}), we show upper limits (red arrows).
We also report the values observed at $z=0$ for spirals and Ultra Luminous InfraRed Galaxies (ULIRG) \citep{kennicutt:1998}, for ULIRGs and Sub-Millimeter Galaxies (SMG) \citep{bothwell2010}, for low-redshift ($1 \lsim z \lsim 2$) galaxies \citep{bouche:2007,tacconi:2010,daddi:2010} and inferred from high-redshift ($z>5.5$) targets \citet[][see the text for details]{vallini:2021}.
The dashed line is the \citet{heiderman:2010} fit for to \citet{kennicutt:1998}, reported in eq. \ref{sk_heiderman}.
As an inset, we add the probability distribution function (PDF) of the burstiness parameter ($\kappa_s$, eq. \ref{ks_deviation}) for galaxies in \code{serra}
\label{fig:schmidt_kennicutt}
}
\end{figure}

The Schmidt-Kennicutt (KS) relation for \code{SERRA} galaxies is plotted in Fig. \ref{fig:schmidt_kennicutt}. We find that simulated galaxies are typically located \textit{above} the local relation \citep{heiderman:2010}
\be\label{sk_heiderman}
\sigmasfr^{\rm KS} = 10^{-12}(\Sigma_{g}/\surfd)^{1.4}\,\surfsfr\,,
\ee
that is a fit to the \citet{kennicutt:1998} observations.

Note that 11 galaxies are temporarily quenched at $z=7.7$, as discussed in the previous Section. Observations of both local starbursts \citep{kennicutt:1998,bothwell2010} and $1 \lsim z \lsim 2$ galaxies \citep{bouche:2007,tacconi:2010,daddi:2010} match our results quite well.

The deviation from the KS can be quantified via the \quotes{burstiness} parameter \citep{ferrara:2019,pallottini:2019}
\be\label{ks_deviation}
\kappa_s \equiv \sigmasfr/\sigmasfr^{\rm KS}\,,
\ee
which also governs the observability of several emission lines \citep{vallini:2020,vallini:2021}.
At $z=7.7$, galaxies in \code{serra} have an average value of $\kappa_s = 3.03^{+4.9}_{-1.8}$.

Unfortunately, CO detections \citep{pavesi:2019,dodorico:2018} in EoR galaxies are rather difficult and rare, making a precise assessment of the gas mass rather uncertain, albeit indirect methods \citep[e.g.][that propose to use \CII~as a proxy for molecular gas]{zanella:2018} might overcome such problem.
An alternative possibility consists of combining \CII~and \CIII~lines to determine $\kappa_s$, the gas density and metallicity \citep[][see also \citealt{ferrara:2019} for details on the physical model]{vallini:2020}.
This method has been applied to COS-3018 ($z=6.854$), the only normal EoR galaxy so far detected in both \CII~\citep[][see \citealt{carniani:2018} for the size determination]{smit:2018} and \CIII~\citep{laporte:2017_b}. \citet{vallini:2020} infer $\kappa_s\simeq 3$ for COS-3018, in line with the average value from \code{serra}.

Using a similar approach, but based instead on the FIR \CII~and \OIII~lines, \citet{vallini:2021} find $\kappa_s \simeq 36.3^{+19.1}_{-18.7}$ for 10 $z>5.5$ galaxies. This value is in line with the values found for \code{serra} galaxies with approximately $\Sigma_{g} \simeq 10^9 \msun {\rm kpc}^{-2}$. 

\subsection{Size-stellar mass relation}\label{sec_size_stellar_mass}

\begin{figure}
\centering
\includegraphics[width=0.49\textwidth]{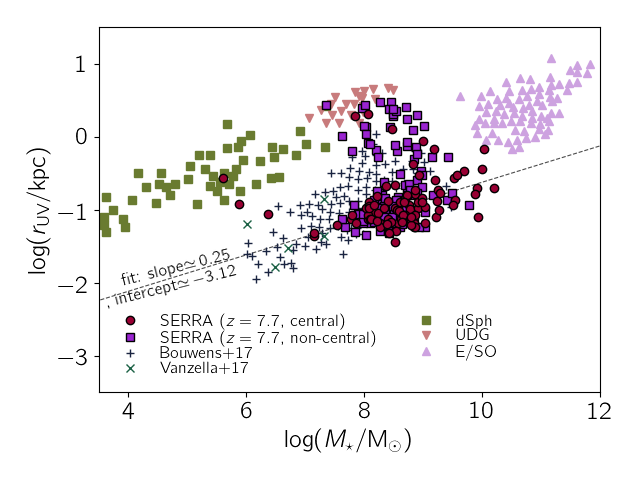}
\caption{
Size-stellar mass relation. The size of \code{serra} galaxies is taken to be equal to $r_{\rm UV}$, the radius encompassing $50\%$ of the UV luminosity. As indicated in the legend, we distinguish between central and non-central galaxies at $z=7.7$, .
As an inset, we overplot the fit for the central galaxies in \code{SERRA}, obtained by excluding the outliers due to mergers (see eq. \ref{fit:size_mass}).
We report the observations for $6\leq z\leq 8$ lensed galaxies \citep{bouwens:2017_b} and \quotes{proto-globular} clusters at $z\gsim 3$ \citep[][\citealt{vanzella:2018}]{vanzella:2017}.
We also show data for a collection of local galaxies \citep{brodie:2011,norris:2014}: dwarf spheroidals (dSphs), Ultra Diffuse Galaxies (UDG), and elliptical/lenticular (E/S0).
\label{fig:size_mass}}
\end{figure}

The size-stellar mass relation for our sample is reported in Fig. \ref{fig:size_mass}.
We define the galaxy size, $r_{\rm UV}$, as the radius that contains $50\%$ of the UV luminosity\footnote{In simulations for which the continuum radiative transfer is not available, it is common to adopt the $50\%$ or $80\%$ stellar mass radius. With respect to $r_{\rm UV}$, this choice underestimates the size by a factor of a few, since UV scattering, that depends on the dust distribution, is not accounted for.}.
In our sample, the behaviour is two folded: most of the galaxies have $r_{\rm UV}\simeq 100\, \rm pc$, while about $10\%$ of the objects have larger sizes ($\simeq 1\, \rm kpc$).
The latter group is composed by galaxies -- mostly non central -- that have recently experienced or are experiencing a merger (e.g. see later in Fig. \ref{fig:zinnia_maps}). While their size-stellar mass relation resembles that of local Ultra Diffuse Galaxies \citep[UDG]{brodie:2011,norris:2014}, this might be a transient feature related to the merging phase. 
Fitting the relation for the central \code{serra} galaxies excluding the mergers, we obtain:
\be
\label{fit:size_mass}
\log \frac{r_{\rm UV}}{\kpc} = (0.249\pm 0.002) \log \frac{M_\star}{\msun} -3.12\pm 0.13
\ee
The size-stellar mass relation from \code{serra} has a slope consistent with the corresponding one in the local universe encompassing dwarf spheroidals (dSphs), Ultra Diffuse Galaxies (UDG), and elliptical/lenticular (E/S0); however in \code{serra} the relation is offsetted by about an order of magnitude, qualitatively as expected from the redshift evolution \citep{shibuya:2015}.

Our data is broadly consistent with that derived by \citet{bouwens:2017_b} for $6\leq z\leq 8$ lensed galaxies and by \citet{vanzella:2018} for $z\gsim 3$ for \quotes{proto-globular} clusters.
However, the median value of the simulated relation is downshifted by a factor of $\simeq 2$ with respect to \citet{bouwens:2017_b} data.
The reasons for this discrepancy are uncertain, but they might be related to the feedback prescriptions adopted in \code{SERRA}. For example, it might be that the energy deposition is insufficient to drive a sufficient amount of gas away from the galactic centre.
While the main goal of all feedback models is to primarily prevent excessive star formation, the resulting kinematic behaviour may considerably differ among them \citep[see][for a in depth study]{rosdahl:2017}. 

For example, a feedback based on delayed-cooling, which shares many of the features of the physically-motivated feedback implemented in \code{SERRA}, tends to puff-up the gas distribution. A kinetic feedback, instead, is known to promote powerful outflows. 
Intriguingly, though, in \citet{lupi:2020} we adopted a kinetic feedback prescription following \citet{hopkins:2018}, and same initial conditions to model the galaxy Freesia presented in \citet{pallottini:2019}. By comparing the two simulations we find a difference of only a few percent in terms of the galaxy size.

Alternatively, part of the tension might be alleviated by properly considering observational uncertainties. As shown in \citet{zanella:2021}, depending on the band and the resolution of the synthetic observation, we can expect that adopting a realistic instrumental noise for the simulated data can give a $\simeq 20\%$ error on the recovery of $r_{\rm UV}$.
This error estimate for the mass-size relation is likely to be a lower limit, since in \citet{zanella:2021} we consider neither lensing model uncertainties nor stellar mass determination errors. Further study on the size-mass relation and a full morphological analysis are left for future work.

\subsection{Emission properties}\label{sec_emission_7_7}

\begin{figure*}
\centering
\includegraphics[width=0.33\textwidth]{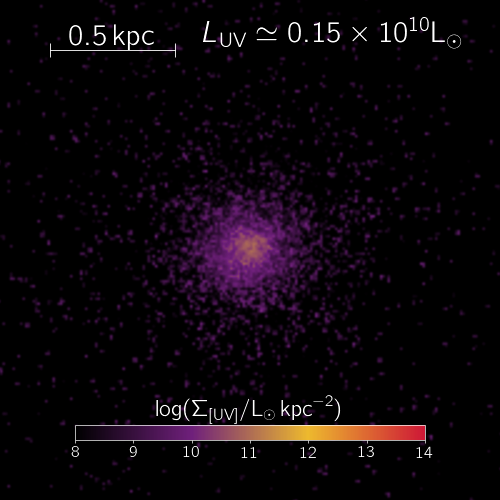}
\includegraphics[width=0.33\textwidth]{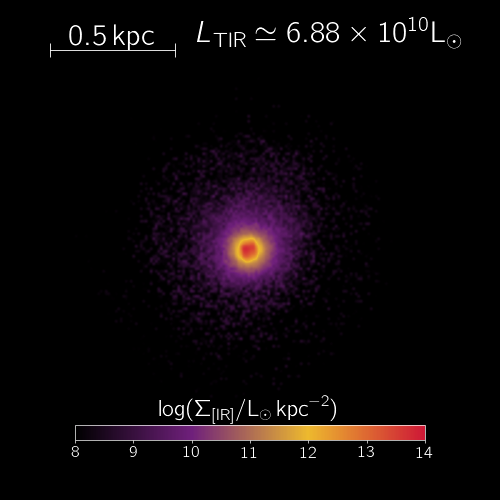}
\includegraphics[width=0.33\textwidth]{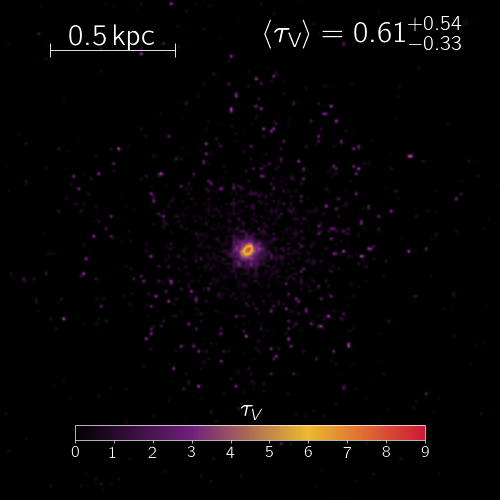}

\includegraphics[width=0.33\textwidth]{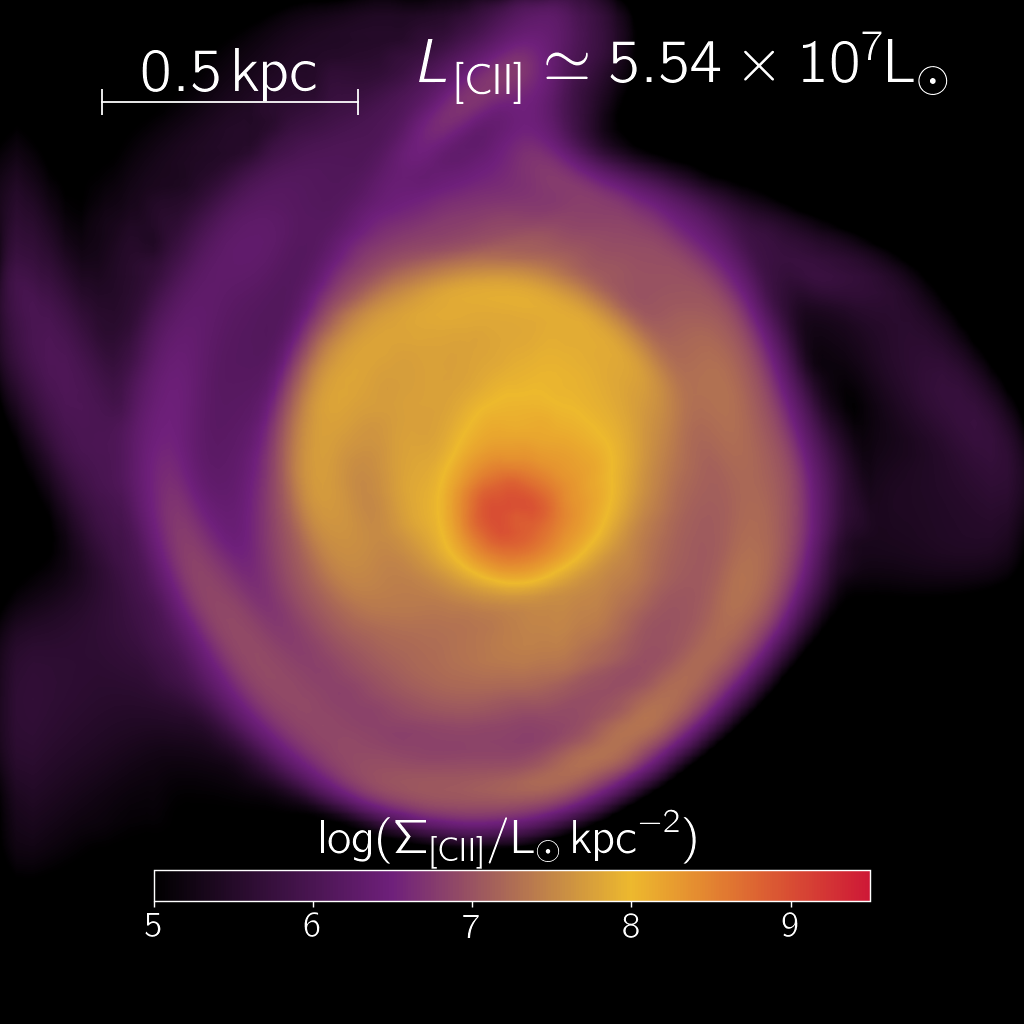}
\includegraphics[width=0.33\textwidth]{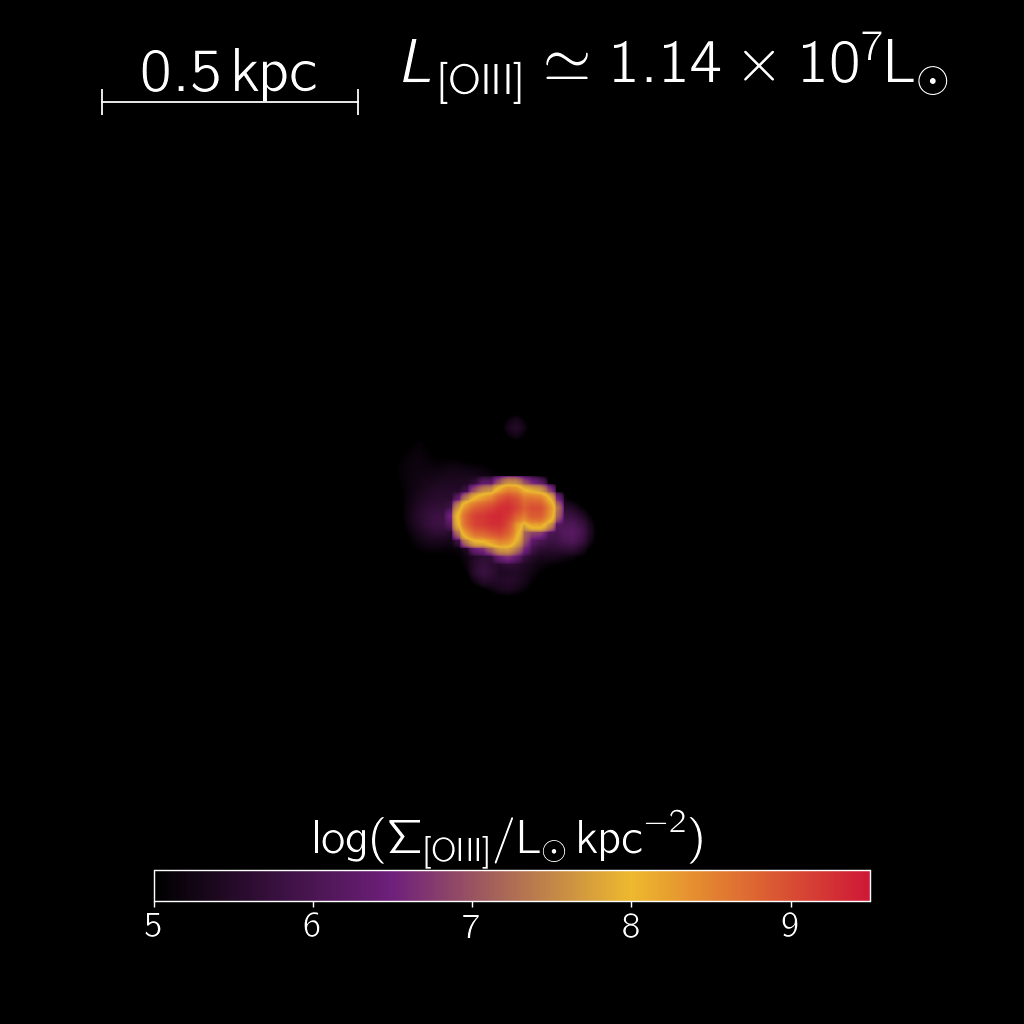}
\includegraphics[width=0.33\textwidth]{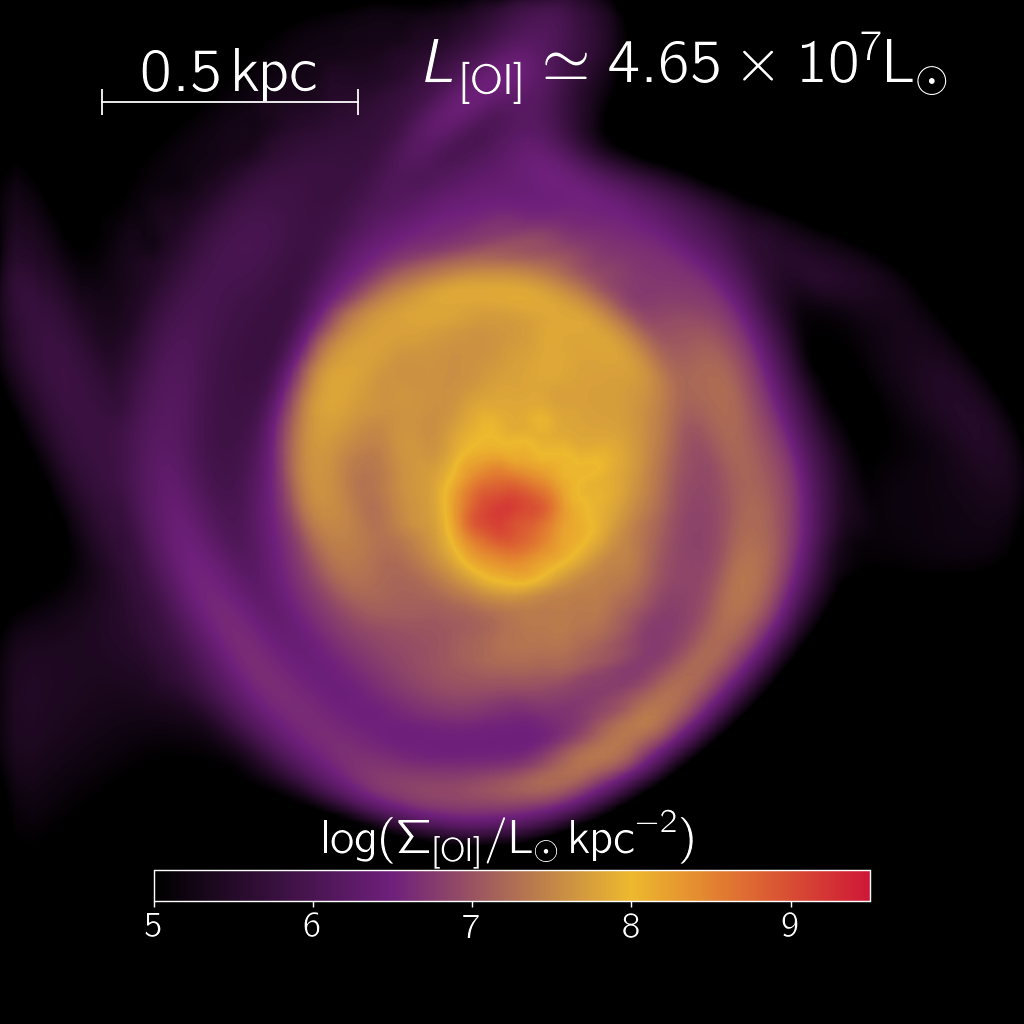}
\caption{
Emission properties of \quotes{Dianthus}, a $M_\star = 2.53 \times 10^9 \msun$ star forming galaxy in \code{SERRA} ($z=7.7$).
The {\bf upper} panels show the continuum properties: the surface brightness in the restframe UV ({\bf left}, $\Sigma_{\rm UV}$, $1000 \leq \lambda/{\rm \AA} \leq 3000$) in the total IR emission ({\bf center}, $\Sigma_{\rm TIR}$, $8 \leq \lambda/\mu{\rm m} \leq 1000$), and the V-band optical depth ({\bf right}, $\tau_V$).
The {\bf lower} panels show the surface brightness of a set of FIR emission lines: \CII~({\bf left}, 158$\mu$m), \OIII~({\bf center}, 88$\mu$m), \OI~({\bf right}, 61$\mu$m).
The maps are taken in a FOV of $2\,\rm kpc$ with a face-on orientation.
As insets, we report the total luminosity of each emission map and the average value of $\tau_V$.
Models for line and continuum emission are described in Sec.s \ref{sec_emission_line} and \ref{sec_emission_continuum}, respectively.
\label{fig:dianthus_maps}
}
\end{figure*}

\begin{figure*}
\centering
\includegraphics[width=0.33\textwidth]{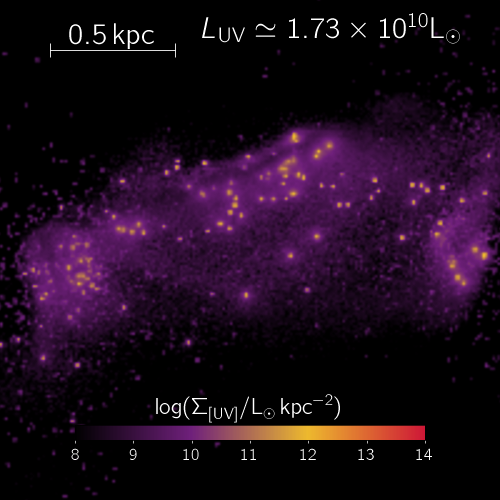}
\includegraphics[width=0.33\textwidth]{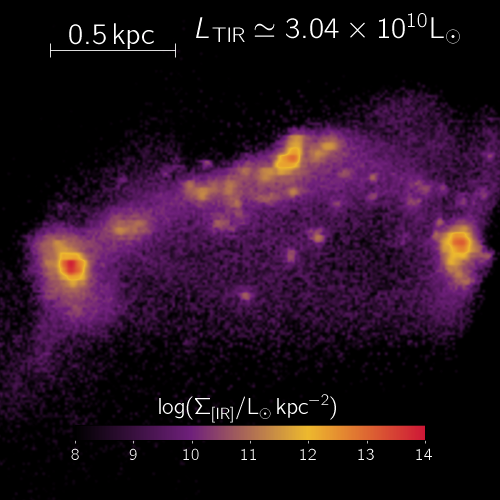}
\includegraphics[width=0.33\textwidth]{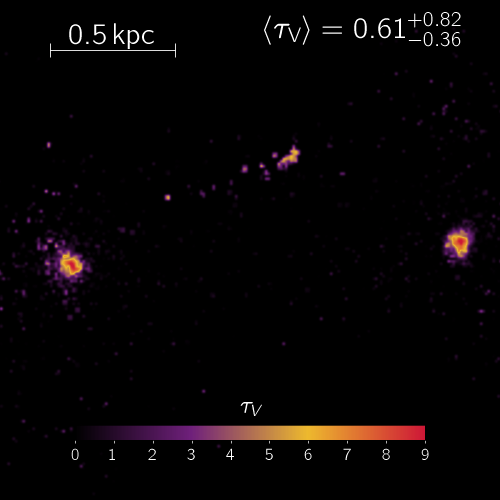}

\includegraphics[width=0.33\textwidth]{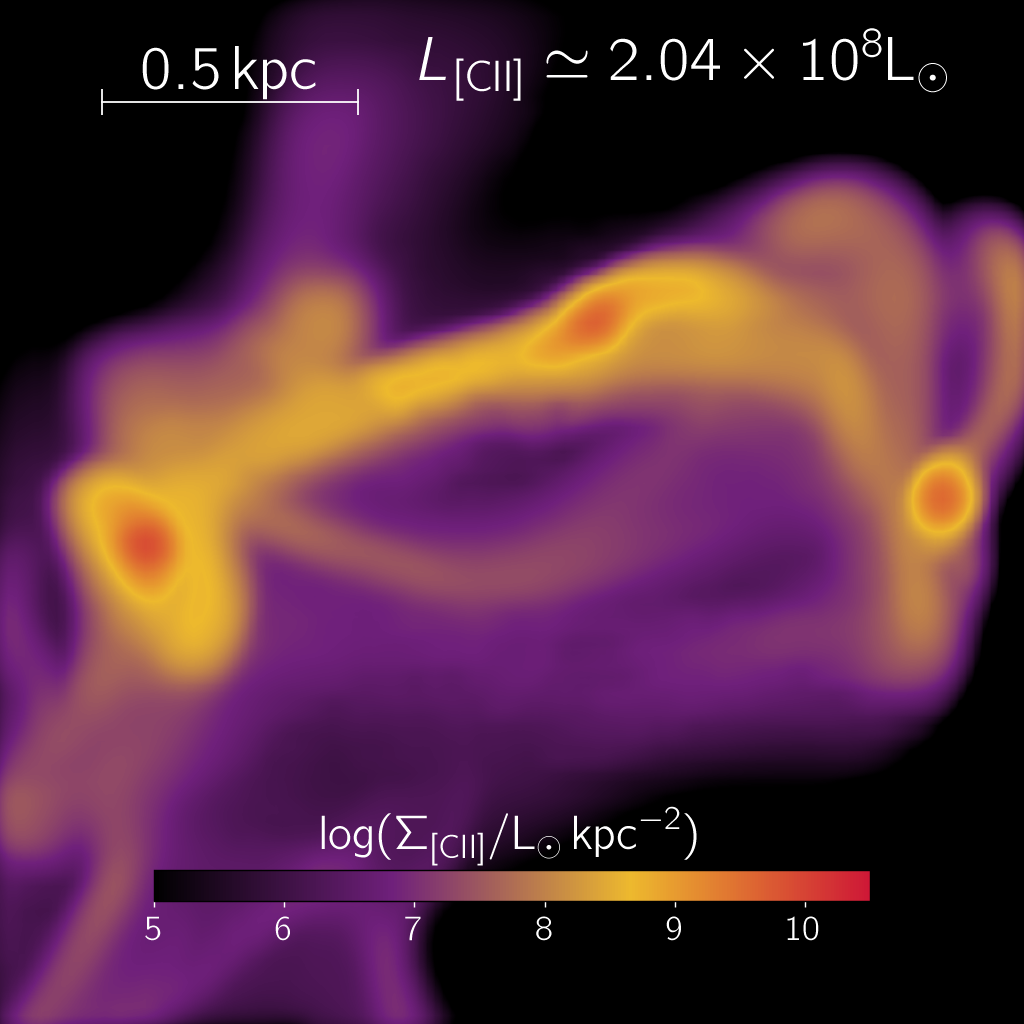}
\includegraphics[width=0.33\textwidth]{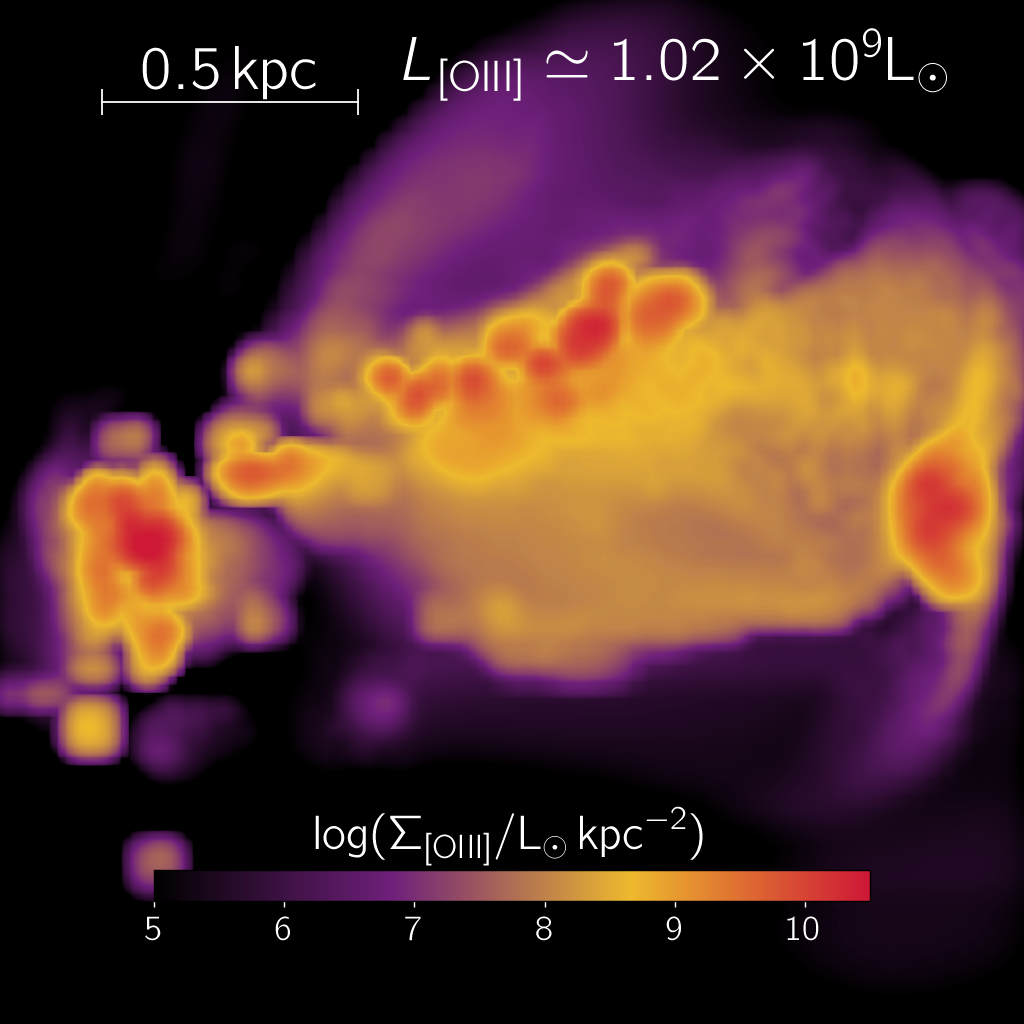}
\includegraphics[width=0.33\textwidth]{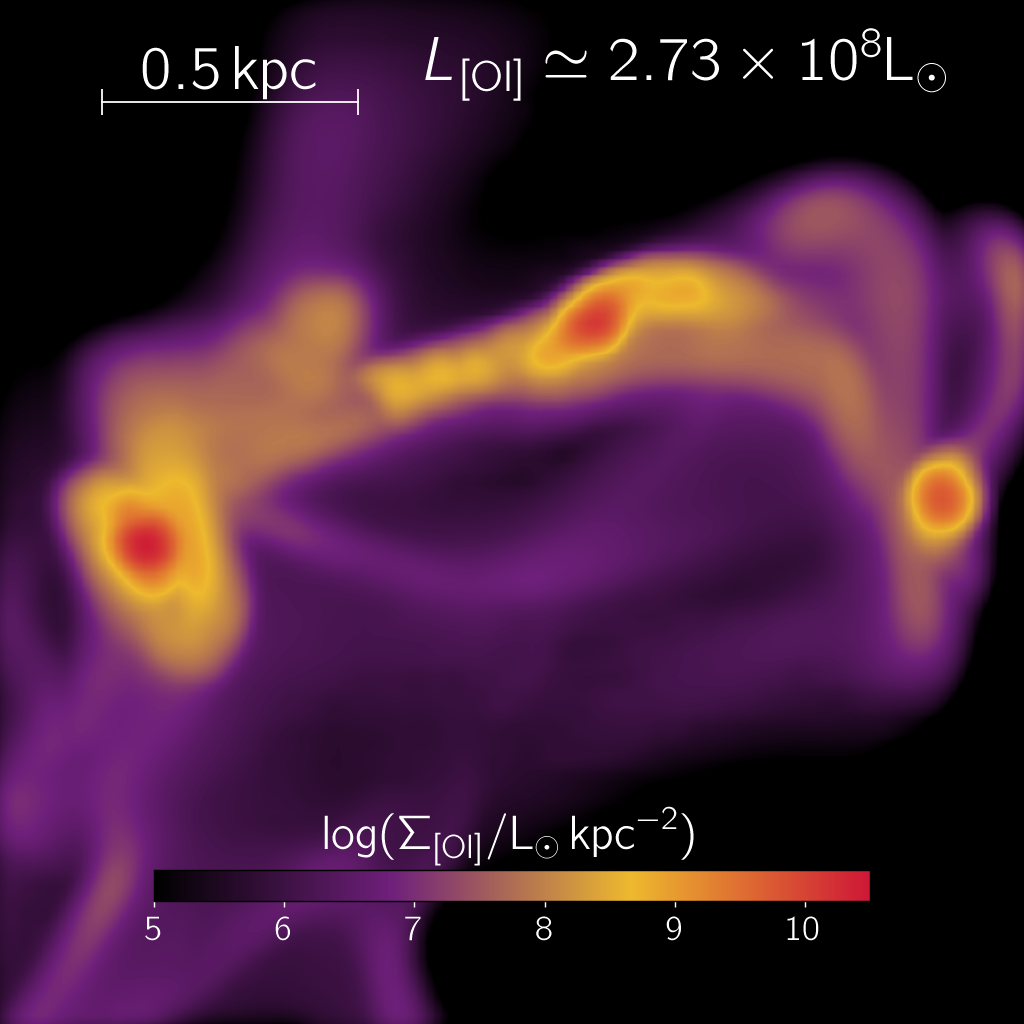}
\caption{
Emission properties of \quotes{Zinnia}, a $M_\star = 2.19 \times 10^9 \msun$ merging galaxy in \code{SERRA} ($z=7.7$). Notation as in Fig. \ref{fig:dianthus_maps}.
\label{fig:zinnia_maps}}
\end{figure*}

Before presenting statistical results for the entire \code{SERRA} sample at $z=7.7$, we want to focus on the specific cases of \texttt{02:46:3788} and \texttt{05:46:6043}, which are also known as \quotes{Dianthus} and \quotes{Zinnia}, respectively.

Both Dianthus and Zinnia have have similar stellar masses ($M_\star = 2.53 \times 10^9 \msun$ and  $M_\star = 2.19 \times 10^9\msun$), star formation rates (${\rm SFR} = 11.13\, \msunyr$ and ${\rm SFR} = 24.46\, \msunyr$) and both are moderately bursty ($\kappa_s = 8.46$ and $\kappa_s = 4.29$), i.e. within about the average of the survey.

The main emission properties of Dianthus and Zinnia are shown in Fig.s \ref{fig:dianthus_maps} and \ref{fig:zinnia_maps}, respectively\footnote{Here we focus on face-on views of the systems, however, we note that morphology and inclination can be important factors in determining the spectral shape and observed properties of galaxies \citep{behrens:2018,katz:2019MNRAS,kohandel:2019}.}.
For the two, the morphology is very different: while Dianthus appears as a regular star forming galaxy, Zinnia has just experienced a merger (a very common situation at these high redshifts), and its stars/gas are still in the relaxation process (see Fig. \ref{fig:warhol_overview}).

Dianthus has a single peak for both continuum and line emission. The stellar component has a radius $\simeq 0.1\,\kpc$, with the UV being slightly less concentrated than the IR continuum. Instead, both the \CII~and the~\OI -- tracing the gas distribution -- have typical radii of $\simeq 0.4\,\kpc$, i.e. with a Perito ratio ($r_{\rm [CII]}/r_{\rm UV}$) similar to what is observed for high-$z$ targets \citep{carniani:2018} but not as extreme as for the \CII~halos observed for galaxies with higher masses \citep{ginolfi:2020,fujimoto:2020}. The \OIII, being linked to the ionization parameter, is much more concentrated.
The morphology of Zinnia appears elongated, with a maximum extension of about 1.5 kpc; three main components are clearly visible in the IR continuum ($\Sigma_{\rm IR} \gsim 10^{12} \surfl$), and in various FIR lines ($\Sigma_{\rm line} \gsim 5\times 10^{9} \surfl$, for \CII, \OI, and \OIII]). The FIR line emission from different ions is mostly co-spatial for Zinnia, with a few noticeable differences.
While \CII~and \OI~trace similar gas phases, the \OI~is more concentrated and shows higher surface brightness in the 3 main components ($\Sigma_{\rm [OI]} \simeq 5\times 10^{10} \surfl$). The \CII~is instead fainter at the peaks ($\Sigma_{\rm [CII]} \simeq 10^{10} \surfl$) but brighter in the outskirts.
The \OIII~emission is particularly intense ($\Sigma_{\rm [OIII]} \simeq 5\times 10^{10} \surfl$) and extended, making it the most luminous line, i.e. $L_{\rm [OIII]} \simeq 5 L_{\rm [CII]} \simeq L_{\rm [OI]}$.
Note that for both galaxies, the luminosity of \CII, \OI, \OIII~is within the observed scatter for local dwarf galaxies with similar SFR \citep[][]{de_looze:2014}.

In Dianthus and Zinnia, all these prominent star forming knots are not visible in the UV, because these regions are significantly obscured ($\tau_V\gsim 5$). On average, though, both galaxies are basically transparent, with $\langle \tau_{V}\rangle \simeq 0.61$. Interestingly, such a large optical depth is caused by a relatively little amount of dust ($M_d\sim 10^5\msun$) distributed in a compact region around the star forming sites. As a result of such geometry, the dust becomes hot.

In particular, the SED-fitted temperature is $T_{\rm SED}\simeq 88\, \rm K$ for Zinnia \citep{sommovigo:2021}. However, the mass-weighted temperature is lower, $\langle T_{\rm d}\rangle_{\rm M} \simeq 61.8 \, \rm K$, $T_{\rm SED}$ is skewed toward the luminosity-weighted temperature of $\langle T_{\rm d}\rangle_{\rm L} \simeq 117.3 \, \rm K$. The latter is boosted by the aforementioned pockets of dust around active star forming regions (already noted in \citealt[][]{behrens:2018}) which therefore produce a blatant MIR excess over a standard grey-body spectrum. We warn that such large temperatures might be partially an artefact of numerical resolution; we defer the discussion of this issue to Sec. \ref{Sec:continuum}.

The emission properties of Zinnia (such as e.g. luminosity and surface brightness ratios, value for the luminosity of different tracers with SFR) are pretty typical of galaxies in \code{serra}. However, the morphological properties (extent of the line emission, multi-component emission) are highly disturbed by the merger episode. For observations of such multi-component galaxies, single-zone assumptions and usual calibrators (e.g. IRX-$\beta$ relation) adopted to infer their properties should be taken with care, as detailed in \citet{ferrara:2022} for the analysis of \code{rebels} galaxies \citep{bouwens:2021}.

In the next Sections we turn to the discussion of the entire galaxy sample.

\subsubsection{UV production}\label{sec_uv_production}

\begin{figure*}
\centering
\includegraphics[width=0.49\textwidth]{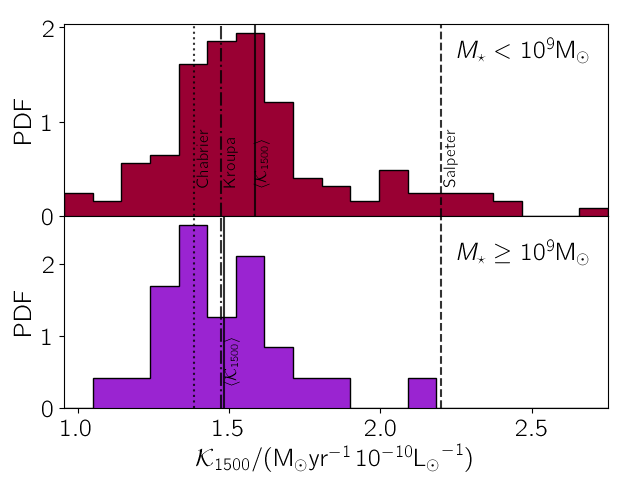}
\includegraphics[width=0.49\textwidth]{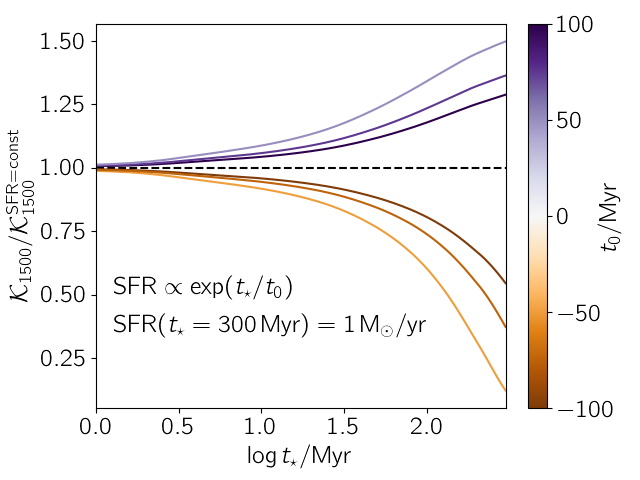}
\caption{UV-SFR conversion factor ($\mathcal{K}_{1500}$, eq. \ref{eq:conversion_uv}).
{\bf Left panel}: distribution of the $\mathcal{K}_{1500}$ in \code{serra}. We separately plot the $\mathcal{K}_{1500}$ PDF for galaxies of low ($M_\star\leq 10^9 \msun$, higher panel) and high ($M_\star\geq 10^9 \msun$ lower panel) stellar mass.
The average of the two subsamples is plotted with a vertical black solid line.
As a reference, we plot as vertical lines the values expected for \textit{constant} star formation histories of stellar population with $Z_\star=\zsun$, age $t_\star>300\,\rm Myr$ in the $0.1-100\,\msun$ range for \citet{salpeter:1955}, \citet{kroupa:2001}, and \citet{chabrier:2003} IMFs.
{\bf Right panel}: variations of $\mathcal{K}_{1500}$ due to different star formation history (SFRh). SFRh are exponential functions parametrised by $t_0$ (see text), as shown in the colorbar. For each SFRh, as a function of the age we plot $\mathcal{K}_{1500}$ normalised by the value for a constant SFRh. All SFRh have ${\rm SFR}=1\,\msunyr$ at $300\, \myr$. For reference, we plot with a dashed line the value for a constant SFRh ($t_0 = \pm\infty$).
\label{fig:k1500}
}
\end{figure*}

To quantify the production of UV photons by \code{SERRA} galaxies we introduce the commonly adopted \citep{madau:2014} conversion factor $\mathcal{K}_{1500} $, defined by 
\be\label{eq:conversion_uv}
{\rm SFR} \equiv \mathcal{K}_{1500}\, L_{1500}\,.
\ee
Observationally, $\mathcal{K}_{1500} $ is a particularly relevant quantity, since it can be used to infer the SFR from the UV data \citep[e.g.][]{kennicutt:2012,madau:2014}. In general, $\mathcal{K}_{1500}$ depends on the IMF of the stellar population, its metallicity, and SFR history (SFRh).
In the left panels of Fig. \ref{fig:k1500} we show the distribution of $\mathcal{K}_{1500}$ for the 202 \code{serra} galaxies, by using the intrinsic (i.e. unattenuated) $L_{1500}$ value and computing the SFR in the last $20\,\myr$. Both distributions for low ($M_\star < 10^9\msun$) and high ($M_\star \geq 10^9\msun$) stellar masses show a central wide peak, while the tail is skewed at high (low) $\mathcal{K}_{1500}$ values for the low (high) masses\footnote{For the PDF $\mathcal{K}_{1500}$ we show the distinction between high and low mass galaxies, since calculating the PDF for central and non-central galaxies gives negligible differences.}.

The average is $\langle \mathcal{K}_{1500}\rangle = 1.53_{-0.13}^{+0.15} ( 1.44_{-0.09}^{+0.14})\times 10^{-10} \msunyr\,\lsun^{-1}$ for the low (high) stellar mass subsample. For the full sample we find instead $\langle \mathcal{K}_{1500} \rangle = 1.52_{-0.14}^{+0.16} \times \,10^{-10} \msunyr\,\lsun^{-1}$.
The above mean values are very close to the predictions ($\mathcal{K}_{1500} = 1.47\times 10^{-10} \msunyr\,\lsun^{-1}$) from a \citet{kroupa:2001} IMF with $Z_\star=\zsun$, stellar age $t_\star>300\,\rm Myr$, and \textit{constant} SFR \citep{madau:2014}. Although this might appear as a trivial result, given that a \citet{kroupa:2001} IMF is also assumed in \code{SERRA}, the situation requires a deeper inspection.

In the right panel of Fig. \ref{fig:k1500}, we show the dependence of $\mathcal{K}_{1500}$ from the SFRh. We calculate $\mathcal{K}_{1500}$ for a \citet{kroupa:2001} IMF with $Z_\star=\zsun$ and selecting an exponential function as a representative history, i.e. ${\rm SFR}(t_\star)\propto\exp(t_\star/t_0)$. We allow increasing and decreasing SFRh by changing the timescale $t_0$: increasing (decreasing) SFRh has a larger (lower) $\mathcal{K}_{1500}$ value with respect to a constant SFR, as newly formed stars weight more in the $\mathcal{K}_{1500}$ computation.
Generally, \code{serra} galaxies feature a time-increasing SFRh (Fig. \ref{fig:time_evolution_galaxy}); however, the increase seen in the range of $\mathcal{K}_{1500}$ is relatively modest ($\simeq 30\%$), particularly for $M_\star \geq 10^9\msun$ galaxies, since their equivalent timescale is $t_0\sim 30\,\myr$ \citep[see Fig. 2 in][]{pallottini:2017_b}.
The frequent bursts experienced by \code{serra} galaxies determine the spread of the $\mathcal{K}_{1500}$ PDF at higher and lower values around the mean, as a consequence of the sudden increase of the SFR and subsequent quenching due to stellar feedback, respectively.
As a note of caution, while the difference between constant and exponentially increasing SFRh is relatively modest because of the expected value for $t_0$ from \code{serra}, assuming a decreasing SFRh for a SED fitting of a galaxy with an increasing SFRh can give a factor $\times 4$ of variation in the SFR determination \citep[see also][]{behrens:2018}.

\subsubsection{Continuum properties}\label{Sec:continuum}

\begin{figure}
\centering
\includegraphics[width=0.49\textwidth]{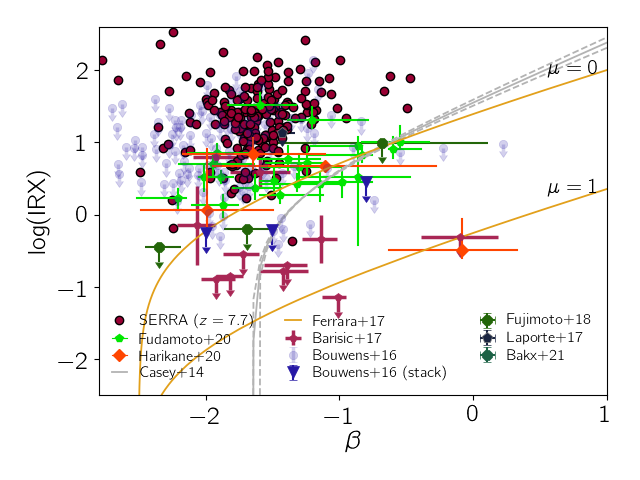}
\caption{
IRX-$\beta$ relation for $z>5$ galaxies compared with \code{serra} galaxies. Data points refer to upper limits \citep{bouwens:2016_b,fujimoto:2017} and detections \citep{laporte:2017,barisic:2017,fudamoto:2020,harikane:2020,bakx:2021}.
Also shown is same relation for local galaxies \citep{casey:2014}, and that expected from a theoretical work \citep{ferrara:2017}. For the latter we show the two limiting cases in which the ISM molecular fraction is either $\mu=0$ (fully atomic) or $\mu=1$ (fully molecular). 
\label{fig:irx_beta}}
\end{figure}
From the spectra of \code{serra} galaxies we compute\footnote{In computing the IRX, we use a UV luminosity integrated in the 1000-3000 \AA\, band. The differences with the often adopted monochromatic luminosity $\lambda_{1600} L_{1600}$ at 1600 \AA\, is within few percent.} ${\rm IRX} \equiv L_{\rm TIR}/L_{\rm UV}$, and the slope of the UV spectrum $\beta$ in the range ($1600-2500$) \AA. The IRX$-\beta$ relation is a convenient summary diagnostic that accounts for the relative balance between UV and FIR photons \citep{calzetti:1994,meurer:1995}. The results are presented in Fig. \ref{fig:irx_beta}, along with high redshift detections and upper limits.

The bulk of \code{serra} galaxies have $-2.5 < \beta < -1$, as expected from a stellar population dominated by relatively unobscured, young stars. However, their IRX is typically $\sim 10$, as expected from evolved and obscured systems.
This apparent contradiction is easily understood as follows. To form stars, the gas must be able to self-shield from the \HH~photo-dissociating UV radiation field. This requires high densities, and the presence of a sufficient amount of dust (see also Fig. \ref{fig:alyssum_hystory}). Fig. \ref{fig:zinnia_maps} shows a clear spatial segregation between regions of high and low $\tau_{V}$. The former have a low volume filling factor, high dust temperatures, and dominate the FIR emission. Low $\tau_{V}$ regions, previously cleared of dust and gas by stellar feedback, determine the $\beta$ slope, and provide most of the UV contribution. Such two-phase configuration, made of dense, opaque clouds and semi-transparent, diffuse medium, produces the observed spatial separation between UV and FIR emission. 

This implies that the IRX-$\beta$ values of our simulated galaxies are located above the relation predicted by models in which molecular clouds are externally illuminated and contain relatively cold dust \citep{ferrara:2017}. Our results are in line with the suggestion by \citet{casey:2014}, i.e. that galaxies with $> 50\,\msunyr$ are dominated at all epochs by short bursts producing young OB stars that are primarily enshrouded in thick dust cocoons. 
The IRX-$\beta$ emerging from \code{serra} seems broadly consistent with most high-$z$ observations \citep{laporte:2017,fudamoto:2020,harikane:2020}. However, we struggle to reproduce the relation when we consider upper limits \citep{bouwens:2016_b,fujimoto:2017}, and some of the detections \citep{barisic:2017}.
The origin of this tension is two folded.

From the observational side, it is very difficult to determine the TIR luminosity. The information is typically recovered from a SED fitting with a grey body spectrum containing 3 parameters, namely (a) the dust mass ($M_d$), (b) the emissivity ($\beta_d$), and (c) the temperature $T_{\rm SED}$.
Typically only few (1-2) continuum bands are available for most high-$z$ targets \citep[but see][]{bakx:2021}, so most determinations are done by assuming a $T_{\rm SED}$ based on low-$z$ estimates and templates \citep[e.g.][see \citealt{sommovigo:2021} for alternatives]{barisic:2017,bethermin:2020}.
Since $L_{\rm TIR} \propto M_d T_{\rm SED}^{4+\beta_d}$ and typically $\beta_d\sim 1.7-2.0$, IRX is extremely sensitive to the estimated or assumed $T_{\rm SED}$ value. 
Observations \citep{schreiber:2018,bakx:2021} and models \citep{sommovigo:2022} indicate that $T_{\rm SED}$ increases towards high redshifts. Thus, the IRX value of the currently available targets might be underestimated, perhaps indicating that at least some of the galaxies in the EoR are already highly obscured \citep{fudamoto:2021}. More robust dust temperature measurements, only made possible through high-frequency ALMA observations (e.g. BAND8/9) should relieve part of the current tension.

From the modelling side, our simulations might fall short in producing low obscuration systems, perhaps due to an insufficient feedback strength. 
In fact, the feedback treatment in \code{serra}, though accurate, is far from being complete, e.g. it lacks an explicit model for clustered SNe \citep{gentry:2017,martizzi:2020}, and cosmic-rays \citep{semenov:2021,rodriguez:2021}, which at least locally are important contributors to the ISM energy budget. Their inclusion might change the emission properties of our galaxies.

In addition, a resolution of $\sim 10\,\pc$ does not allow us to resolve the internal structure of molecular clouds, where most of the simulated FIR is produced.
The internal gas/dust distribution in a MC can dramatically affect the dust temperature, as shown by \citet{sommovigo:2020}. For example, a uniform distribution might result in dust as hot as $\langle T_{\rm d}\rangle_{\rm L} \simeq 120 \, \rm K$; when a more physical profile accounting for the effects of radiation pressure is adopted, the mean temperature drops to $\langle T_{\rm d}\rangle_{\rm L} \simeq 55 \, \rm K$. Modeling the internal structure of MCs is beyond the reach of even the most refined cosmological zoom-in simulations. 

Finally, simulations of individual MCs shows that -- depending on the metallicity, turbulence and the observed band -- 15\% to 70\% of the photons can escape from the cloud before it is disrupted \citep{decataldo:2020,kimm:2021}. If true, molecular clouds would contribute significantly to UV emission, at the same time reducing the FIR emission from our galaxies. Further exploration of this aspect is ongoing.

\subsubsection{Line emission properties}\label{sec_line_emission}

\begin{figure}
\centering
\includegraphics[width=0.49\textwidth]{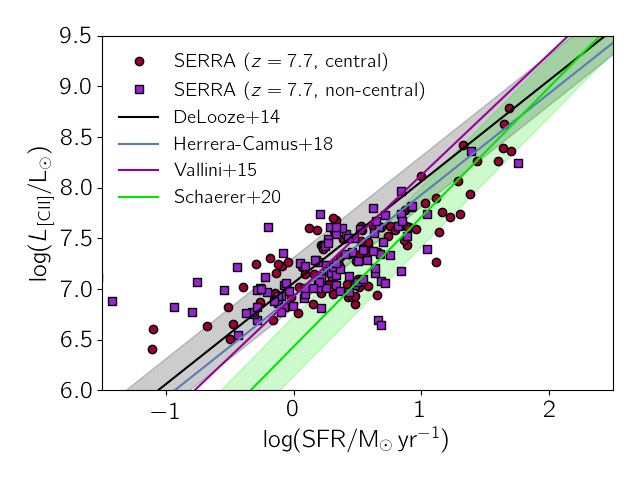}
\caption{
The integrated \CII-SFR relation found in \code{serra}.
Along with \code{serra} data at $z=7.7$ divided in the central and non-central sample, we show the results obtained from local observations \citep{de_looze:2014,herrera-camus:2018}, the relation found at $z=4-6$ by \code{ALPINE} \citep[${\rm SFR}\gsim 10\, \msunyr$][]{schaerer:2020}, and the results from physical motivated models \citep[][$Z=0.5\,\zsun$]{vallini:2015}.
\label{fig:cii_sfr}}
\end{figure}

\begin{figure}
\centering
\includegraphics[width=0.49\textwidth]{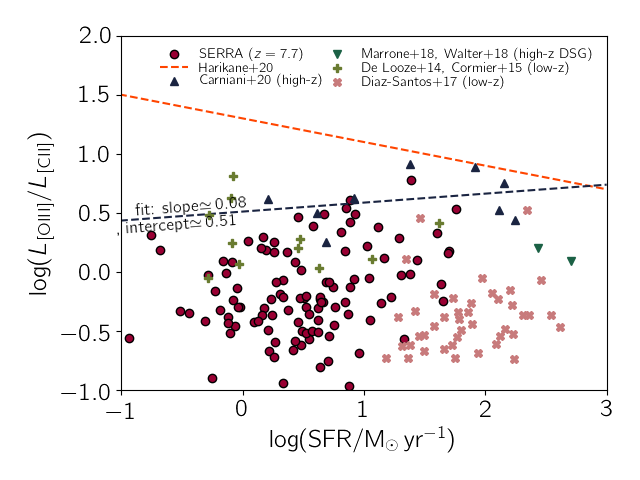}
\caption{~\OIII~to \CII~ratio as a function of star formation rate.
We plot the collection of observations obtained for normal high-$z$ galaxies \citep[$\SFR<100\,\msunyr$][]{carniani:2020} and dusty star forming galaxies \citep[$\SFR>100\,\msunyr$]{walter:2018,marrone:2018}, along with data for local normal \citep{diaz_santos:2017} and dwarfs \citep{de_looze:2014,cormier:2015} galaxies. We also plot the fits from \citet{harikane:2020} and \citet[][computed in the present, best fit parameters indicated in the figure]{carniani:2020}.
\label{fig:oiii_vs_cii}}
\end{figure}

\begin{figure*}
\centering
\includegraphics[width=0.98\textwidth]{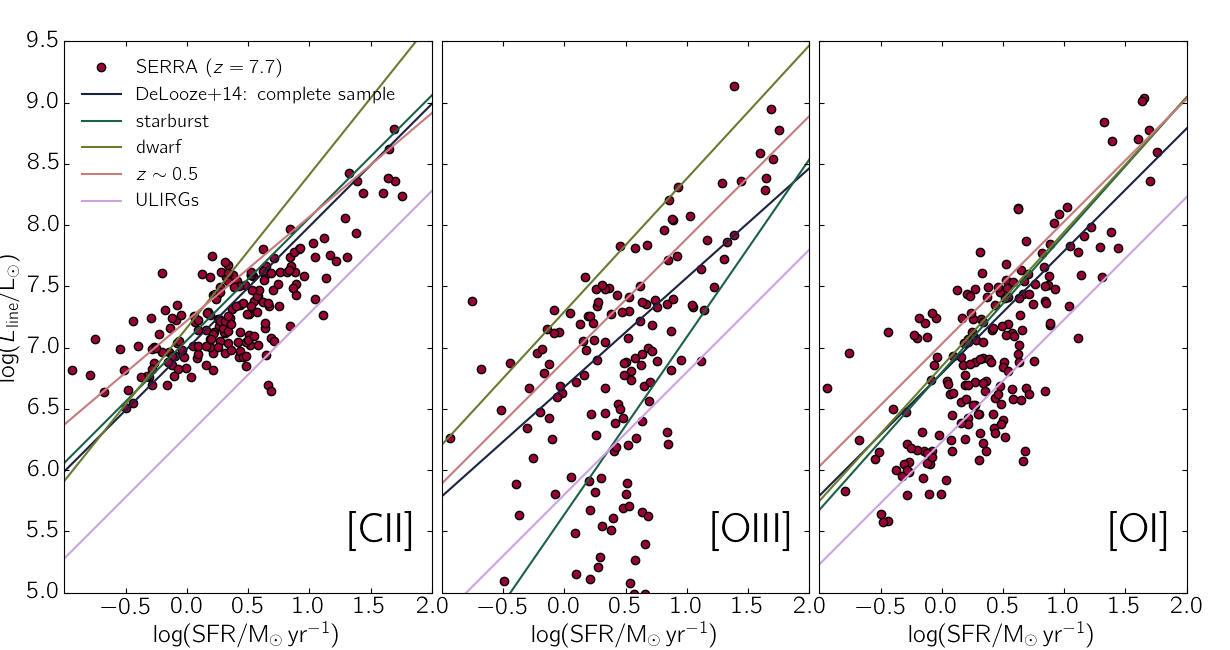}
\caption{
\code{serra} emission line predictions compared to available local data.
We show \CII, \OIII~and \OI~as a function of SFR in the left, central, and right panel, respectively.
Fits to observed trends for different classes of galaxies in the low-$z$ Universe are shown.
The dwarf galaxy observations are taken from \citet{de_looze:2014}, as well as the classification and the relative fits, which are based on additional data from \citet{brauher:2008,graciacarpo:2011,sargsyan:2012,parkin:2013,diazsantos:2013,farrah:2013}.
\label{fig:sfr_indicators}
}
\end{figure*}

The (integrated) \CII-SFR relation is presented in Fig. \ref{fig:cii_sfr}.
Overall, \code{serra} galaxies at $z=7.7$ are consistent with local data \citep{de_looze:2014,herrera-camus:2018}, albeit with a larger scatter. This is expected on the basis of the early work by \citet[][]{carniani:2018} analysing 20 targets, and of the more recent analysis of the \code{ALPINE} sample presented in \citet[][118 targets]{schaerer:2020}.
The scatter in the relation is driven by the larger spread of physical properties (such as $n$, $Z$, $\kappa_s$) deduced for high-$z$ galaxies \citep[][]{vallini:2015,ferrara:2019}; further, the scatter appears to be larger for non-central objects, as a consequence of the interactions with their nearby more massive galaxy.

Overall, \code{serra} galaxies fall onto the \code{ALPINE} relation, but some tension seems present at $\rm SFR \lsim 1\, \msunyr$, where the simulated data seems to flatten and join the local \citet[][]{de_looze:2014} relation. To this concern, we note that the detection limit for \code{ALPINE} galaxies is about $\rm SFR \gsim 10\, \msunyr$ \citep{le_fevre:2019,schaerer:2020}. Deeper ALMA observations would be needed to probe the \CII-SFR at more feeble star formation levels. Additionally, for \code{serra} galaxies at $z=7.7$, the \CII~trend at $\rm SFR \lsim 1\, \msunyr$ shows an offset: small central galaxies tend to have lower \CII~with respect to non-central ones with the same SFR. This is mostly due to the fact that non-central are contaminated, i.e. they are formed in an environment that is pre-enriched by their more massive companion.

While the median values of integrated \CII-SFR presents little differences at high and low redshift, recall that \code{serra} galaxies are typically located below the resolved $\Sigma_{\rm [CII]}$ - $\Sigma_{\rm SFR}$ relation, as also found by high-$z$ observations \citep{carniani:2018}. Such \CII-deficiency has been discussed for a sub-sample of \code{serra} galaxies in \citet{pallottini:2019} and \citet{vallini:2021}. It has been interpreted as a signature of an ongoing bursty star formation activity in these galaxies \citep[high $\kappa_s$][see Fig. \ref{fig:schmidt_kennicutt}]{ferrara:2019}.

Because of the increasing availability of \OIII~$88\mu$m line observations, the \CII/\OIII~ratio has become a viable tool to study the high-$z$ ISM. In Fig. \ref{fig:oiii_vs_cii} we plot the predicted ratio for the \code{serra} sample as a function of $\rm SFR$, along with an updated collection of both high-$z$ and local observations.
The $L_{\rm [OIII]}/L_{\rm [CII]}$ ratio for \code{serra} galaxies is in the range $0.1-10$, and shows only a mild correlation with $\rm SFR$. A non negligible fraction of our systems has a low $L_{\rm [OIII]}/L_{\rm [CII]}$, as a result of the intermittent star formation activity discussed in Sec. \ref{Sec:Alyssum}. If the SFR is quenched for more than $\simeq 20\,\rm Myr$, the production of \OIIIion~ionising photons drops, and the \OIII~emission consequently fades away.
This effect might be overestimated in \code{serra}, since in principle a better mass resolution would yield a finer sampling of the SFR process in those cases where the stellar feedback suppresses the SFR without completely quenching it. However, even simulations with a better mass resolution \citep[e.g.][$\simeq 4\,\msun$ vs $\simeq 1.2\times 10^4 \msun$ adopted here]{gutcke2021} show a complete quenching of the SFR activity for long periods of time in low mass ($M_\star\lsim 10^9\msun$) systems.

Moreover, the observed $L_{\rm [OIII]}/L_{\rm [CII]}$ values \citep{harikane:2020} must be down revised when accounting for the concentration bias \citep{carniani:2020}, i.e. since \CII~is more extended then \OIII, its low surface brightness emission is easier to miss, and implies that $L_{\rm [CII]}$ can be underestimated by a factor of $\simeq 2$, depending on the angular resolution and signal to noise ratio of the observation.

Fitting the data from \citet{carniani:2020} we obtain
\be
\log \frac{L_{\rm [OIII]}}{L_{\rm [CII]}} = (0.08\pm 0.01) \log \frac{\rm SFR}{\msunyr} + 0.51\pm 0.02\,
\ee
i.e. a flatter curve with respect to \citet{harikane:2020}, with rations that are about one order of magnitude lower at ${\rm SFR}\sim 1\msunyr$, where we currently have very few detections. Indeed, some of our galaxies can match the observed rations, however the bulk of \code{SERRA} galaxies have $L_{\rm [OIII]}/L_{\rm [CII]}$ that are lower than the observed ones.
It is interesting to note that while Dianthus (Fig. \ref{fig:dianthus_maps}) and Zinnia (Fig. \ref{fig:zinnia_maps}) have very similar properties ($M_\star$, $\rm SFR$, $\kappa_s$, $\dots$), $L_{\rm [OIII]}/L_{\rm [CII]}$ is $\simeq 0.2$ and $\simeq 5.0$, respectively; the former galaxy is a disk, the latter is a merger, which hints at the possibility that the dynamical state of the system can play an important role in the ratio determination; we are planning to explore the aspect more systematically in the future \citep[see][]{rizzo:2022}.

While we need more observations to robustly determine a relation at high-$z$, this is an indication that the ratios for high-$z$ targets tend to be more in line with those obtained for local galaxies \citep{de_looze:2014,cormier:2015,diaz_santos:2017}, and the \code{serra} ones \citep[see also][]{vallini:2021}. Thus, we find no need to invoke alternative explanations, such as a top-heavy IMF or differential enrichment patterns for C and O, as explored in \citet{arata:2020} and \citet{katz:2021}, which adopt the un-revised values for the ratio.
While we can expect that variations of C and O abundances are present in the ISM of high-$z$ galaxies -- which also have other implications and should be explored -- the current data does not seem to call for extreme scenarios strongly.
We stress that the above discussion must be taken with a grain of salt, as currently only 11 targets at high-$z$ have both \CII~and \OIII~detections; in particular, it is still unclear if we are picking the brightest, more concentrated, \OIII~emitter or seeing the bulk of the population. Considering that observational evidence and models agree on a very mild evolution of $L_{\rm [CII]}-\rm SFR$ with $z$, the latter option would imply that an evolution of the $L_{\rm [OIII]}-\rm SFR$ that is larger than predicted by the simulations, which would entail a revision of part of our modelling: only future observations will be able to test the different scenarios.

To conclude, we analyse the SFR dependence of different lines (\CII, \OIII, and \OI~$63\mu$m) for simulated galaxies and compare it with all the available low-redshift data to assess whether early systems differ substantially from their local counterparts in this respect. In Fig. \ref{fig:sfr_indicators} we show the results using a collection of diverse samples, i.e. starbursts, dwarfs, $z\sim 0.5$ and ULIRGs \citep{de_looze:2014}. 

Regarding the \CII, the bulk of \code{serra} follows the starbursts and $z\lsim 0.5$ relation; interestingly at ${\rm SFR} \gsim 3\, \msunyr$ the simulated galaxies are intermediate between starbursts and ULIRGs, and far from the slope characterising the dwarf galaxy population. Qualitatively, this is expected considering that \code{serra} galaxies have high IRX values for their $\beta$ (Fig. \ref{fig:irx_beta}), a feature shared by many high-$z$ observed targets. Also, some of the \code{rebels} galaxies appear as scaled-down versions of dusty star forming galaxies \citep{fudamoto:2021}.

The \OIII-SFR relation is more complex. Most of the simulated high-$z$ galaxies follow the same trend of $z\lsim 0.5$ systems. However, a consistent fraction of the sample is found along the starburst locus; the remaining part contains galaxies which have their SFR temporarily quenched, and therefore showing a strongly reduced \OIII~luminosity. The presence of single trend for \CII~and two trends for \OIII~is the reason why no $L_{\rm [OIII]}/L_{\rm [CII]}$-SFR relation can be identified (see Fig. \ref{fig:oiii_vs_cii}). The quenching is not fully captured by the value of the sSFR: also the burstiness $\kappa_s$ plays a role, along with mean gas density and metallicity, which can give secondary dependences on the \OIII~line \citep{vallini:2021}; the situation is complex, also considering that consistency with local line luminosity-SFR scaling relations (Fig. \ref{fig:sfr_indicators}) does not automatically imply consistency of the line ratio vs SFR scaling (Fig. \ref{fig:oiii_vs_cii}); we aim at addressing this point in a following work.

Finally, \OI~is an important tracer of the dense atomic and molecular regions, but so far it has been detected only in a single $z\simeq 6$ lensed galaxy with APEX \citep{rybak:2020}. Similarly to \CII, \OI~predictions from \code{serra} galaxies follow the trend for starbursts and $z\lsim 0.5$ targets, even tough the scatter appears somewhat larger.
Although \OI~$63\mu$m is more luminous than \CII, unfortunately it falls in ALMA BAND9 for a $z\simeq 6$ object. Let us focus on a ${\rm SFR} \simeq 50\, \msunyr$ galaxy with a FWHM of $\simeq 200\,\kms$ \citep{kohandel:2019}; such object has $L_{\rm OI63\mu m}\simeq 5 \times 10^8\lsun$, thus detection (S/N=5) in BAND9 would require $\simeq 6\, {\rm hr}$ of observing time\footnote{This estimate is critically dependent on the galaxy location with respect to the atmospheric absorption lines, that are abundant in BAND9.}.

Note that \OI~$145\mu$m is shifted in the more favourable ALMA BAND7; the trend with SFR of the luminosity of \OI~$145\mu$m is similar to \OI~$63\mu$m \citep[see also][]{lupi:2020}, however \OI~$145\mu$m is usually $\sim 10$ times fainter, thus a detection would require $\simeq 7\, {\rm hr}$ of observing time, i.e. similar to the other line.

Albeit challenging, detecting \OI~is feasible and crucial to clarify, for example, whether a differential C/O enrichment is in place at high-$z$. While such an experiment can be attempted by using \CII~and \OIII \citep{arata:2020,harikane:2020}, the latter is strongly dependent on ionising radiation, thus \OI~would represent a independent and cleaner probe.
Further, \OI~lines fall both in BAND7 and BAND9; while trying to use the latter is typically regarded as a risky strategy, observations in BAND9 could simultaneously give both the \OI~and crucial information to constraint the dust temperature \citep{bakx:2020}, as it is closer to the continuum peak.

\section{Summary}\label{sec_conclusions}

We introduce \code{SERRA}, a suite of zoom-in high-resolution ($1.2\times 10^4 \msun$, $\simeq 25\,\pc$ at $z=7.7$) cosmological simulations including non-equilibrium chemistry and on-the-fly radiative transfer. The outputs are post-processed to derive galaxy UV+FIR continuum and emission line properties. Results are compared with available multi-wavelength data to constrain the physical properties (e.g., star formation rates, stellar/gas/dust mass, metallicity) of high-redshift $6 \lsim z \lsim 15$ galaxies.

In this flagship paper, we have focused our attention on the $z=7.7$ sample, consisting of 202 targets with a stellar mass range $10^7 \msun \lsim M_\star \lsim 5\times 10^{10}\msun$. These objects are resolved with a mass (spatial) resolution of $1.2\times 10^4 \msun$ ($\simeq 25\,{\rm pc}$). The main highlights are the following.

\begin{itemize}
\item All \code{serra} galaxies show a time-increasing SFR featuring rapid fluctuations, particularly evident at low masses, caused by stellar feedback and merging episodes. The specific SFR ranges from ${\rm sSFR} \sim 100\,\gyr^{-1}$ for young ($t_\star \lsim 100\,\myr$), small galaxies ($M_\star \lsim 10^8\msun$) to ${\rm sSFR} \sim 10\,\gyr^{-1}$ for older ($t_\star \gsim 200\,\myr$) more massive ones ($M_\star \gsim 10^9\msun$), in good agreement with high-$z$ data \citep{jiang:2013,rinaldi:2021}.

\item The simulated stellar-to-halo mass relation is consistent with abundance matching works \citep[][]{behroozi:2013} for large halo masses, i.e. $M_{\rm h}\gsim 5\times 10^{10}\msun$. In smaller halos ($M_{\rm h}\sim 10^{9}\msun$), the intense radiation from the first few formation events easily dissociate $H_2$, causing relatively extended periods during which star formation is completely quenched, and the galaxy stays well below the \citet[][]{behroozi:2013} relation. After a series of bursts, a sufficient amount of dust is produced to allow an efficient self-shielding of \HH, and a continuous SF activity which eventually brings the galaxy onto the stellar-to-halo mass relation.

\item On average \code{serra} galaxies are \quotes{bursty}, i.e. they are located above the Schmidt-Kennicutt relation, with a burstiness parameter (eq. \ref{ks_deviation}) $\kappa_s = 3.03^{+4.9}_{-1.8}$. For galaxies with $\sigmasfr\gsim 10 \surfsfr$, $\kappa_s$ is higher and in agreement with \citet{vallini:2021}, who infer $\kappa_s\simeq 36.3$ for 11 observed high-$z$ galaxies.

\item The size-stellar mass relation from \code{serra} is $\log(r_{\rm UV}/\kpc) \simeq 0.25 \log(M_\star/\msun) -3.1$ (see eq. \ref{fit:size_mass}). The relation has the same slope as the one observed in the local Universe \citep{brodie:2011,norris:2014}, but it is downshifted by about one order of magnitude because of redshift evolution \citep[cfr][]{shibuya:2015}. The data from \code{serra} is consistent with lensed galaxies observed at high-$z$ \citep{bouwens:2017_b,vanzella:2018}, but our median has a systematic offset of a factor of $\simeq 2$.

\item The bulk of \code{serra} galaxies have $-2.5 < \beta < -1$, as expected from a stellar population dominated by relatively unobscured, young stars. However, their IRX is typically $\sim 10$, as expected from evolved and obscured systems. The origin of this apparent contradiction resides in the multi-phase ISM structure in these systems, consisting of IR-emitting molecular clumps embedded in a semi-transparent, UV-emitting diffuse component. This configuration also produces a UV vs. dust continuum spatial offset.

\item Regarding line emission, \code{serra} is consistent with the observed \CII$- \rm SFR$ observed at high-$z$ \citep{schaerer:2020} and with the bulk of the inferred $L_{\rm [OIII]}/L_{\rm [CII]}$, when the observational concentration bias is accounted for \citep{carniani:2020}. Further, our results suggest that the dynamical state of the system (merger vs disk) can play an important role in determining the $L_{\rm [OIII]}/L_{\rm [CII]}$ ratio.

\item Detection of \OI~is feasible but challenging, i.e. for a ${\rm SFR}\simeq 50\,\msunyr$ galaxy about $\simeq 6 (7) \rm hr$ in ALMA BAND9 (BAND7) are needed for the $63\mu$m ($145\mu$m) line. Its detection will bring crucial information on the metal enrichment patterns of individual elements in early galaxy systems. While risky, observations in BAND9 would be particularly rewarding, as they could simultaneously give both the \OI~and crucial information to constraint the dust temperature, as it is closer to the continuum peak.

\end{itemize}

\section*{Acknowledgements}

AP, AF, MK, SC, LV, LS acknowledge support from the ERC Advanced Grant INTERSTELLAR H2020/740120 (PI: Ferrara).
SS acknowledges support from the ERC Starting Grant NEFERTITI H2020/804240 (PI: Salvadori).
Any dissemination of results must indicate that it reflects only the author's view and that the Commission is not responsible for any use that may be made of the information it contains.
This research was supported by the Munich Institute for Astro- and Particle Physics (MIAPP) of the DFG cluster of excellence \quotes{Origin and Structure of the Universe}. Partial support from the Carl Friedrich von Siemens-Forschungspreis der Alexander von Humboldt-Stiftung Research Award is kindly acknowledged.
We acknowledge the CINECA award under the ISCRA initiative, for the availability of high performance computing resources and support from the Class B project SERRA HP10BPUZ8F (PI: Pallottini).
We gratefully acknowledge computational resources of the Center for High Performance Computing (CHPC) at SNS.
We acknowledge usage of the Python programming language \citep{python2,python3}, Astropy \citep{astropy}, Cython \citep{cython}, Matplotlib \citep{matplotlib}, NumPy \citep{numpy}, \code{pynbody} \citep{pynbody}, and SciPy \citep{scipy}.

\subsection*{Data availability}

The derived data generated in this research will be shared on reasonable requests to the corresponding author.
Part of the data used for this study is available at the website \url{http://cosmology.sns.it/data_access.html}.

\bibliographystyle{mnras}
\bibliography{master,codes}

\bsp	
\label{lastpage}
\end{document}